\newcommand{\bra}[1]{\langle {#1} |}
\newcommand{\ket}[1]{| {#1} \rangle}
\newcommand{\inproduct}[2]{\langle #1 | #2 \rangle}

\documentclass[12pt,dvipdfmx]{iopart}

\usepackage{iopams}  
\usepackage[dvipdfmx]{graphicx}
\usepackage{wrapfig}  
\usepackage[dvipdfmx]{color}
\begin{document}

\title[Quantal rotation and its coupling to intrinsic motion in nuclei]{
Quantal rotation and its coupling to intrinsic motion in nuclei}

\author{Takashi Nakatsukasa$^{1,2,3}$, Kenichi Matsuyanagi$^{3,4}$,
Masayuki Matsuzaki$^5$, Yoshifumi R. Shimizu$^6$}

\address{$^1$ Center for Computational Sciences, University of Tsukuba,
Tsukuba 305-8577, Japan}
\address{$^2$ Faculty of Pure and Applied Sciences, University of Tsukuba,
Tsukuba 305-8571, Japan}
\address{$^3$ RIKEN Nishina Center, Wako 351-0198, Japan}
\address{$^4$ Yukawa Institute for Theoretical Physics, Kyoto University,
Kyoto 606-8502, Japan}
\address{$^5$ Department of Physics, Fukuoka University of Education, Munakata, Fukuoka 811-4192, Japan}
\address{$^6$ 
Department of Physics, Graduate School of Science, Kyushu University, Fukuoka 
819-0395, Japan
}
\vspace{10pt}
\begin{indented}
\item[]February 2014
\end{indented}

\begin{abstract}
Symmetry breaking is an importance concept 
in nuclear physics and other fields of physics.
Self-consistent coupling between the mean-field
potential and the single-particle motion is a key ingredient in
the unified model of Bohr and Mottelson,
which could lead to a deformed nucleus
as a consequence of spontaneous breaking of the rotational symmetry.
Some remarks on the finite-size quantum effects are given.
In finite nuclei,
the deformation inevitably introduces the rotation as a
symmetry-restoring collective motion (Anderson-Nambu-Goldstone mode),
and the rotation affects the intrinsic motion.
In order to investigate the interplay between the rotational
and intrinsic motions in a variety of
collective phenomena,
we use the cranking prescription together with the quasiparticle
random phase approximation.
At low spin, the coupling effect can be seen in
the generalized intensity relation.
A feasible quantization of the cranking model is presented,
which provides a microscopic approach to the higher-order intensity relation.
At high spin, the semiclassical cranking prescription works well.
We discuss properties of collective vibrational motions 
under rapid rotation and/or large deformation.
The superdeformed shell structure plays a key role in emergence of a new
soft mode which could
lead to instability toward the $K^\pi=1^-$ octupole shape.
A wobbling mode of excitation, which is a clear signature of the
triaxiality,
is discussed in terms of a microscopic point of view.
A crucial role played by the quasiparticle alignment is presented.
\end{abstract}

%
%
%
%
%

\section{Introduction}
\label{sec:intro}

There are a variety of rotating objects in the universe.
We are living on the rotating earth which is revolving around the sun.
The sun is a part of our rotating galaxy.
The neutron star is often observed as a ``pulsar'' whose rotational period
can be as small as $10^{-3}$ s.
However, if we look into a microscopic world, we find
much faster rotating objects, such as nuclei.
The nuclear rotational period in heavy nuclei
is typically $\tau_{\rm rot}=10^{-19}\sim 10^{-20}$ s.
This time scale is  $100-1,000$ times larger than 
the period of the single-particle Fermi motion inside the nucleus,
$\tau_F\sim 10^{-22}$ s.
Thus, the rotational motion can be treated as ``slow'' motion
at low-spin states.
However, in high-spin states produced by the fusion reaction,
it could reach $\tau=10^{-21}\sim 10^{-22}$ s which is comparable to $\tau_F$.
Therefore, the nuclei provide a unique laboratory to study
rapidly rotating quantum systems under strong Coriolis and
centrifugal fields.

The nucleus is a finite quantum many-body system.
Since the Hamiltonian is rotationally invariant,
its energy eigenstate has a definite total angular momentum $I$.
In order to realize the nuclear rotation, 
the nucleus needs to define its orientation.
Since it is impossible to do it for spherical systems,
a deformed {\it intrinsic} state, which is
produced by  breaking the rotational symmetry, is necessary.
The word ``intrinsic'' means the degrees of freedom approximately
independent of the rotational motion.
The spontaneous breaking of symmetry (SBS) is an important concept
to constitute the unified model of Bohr and Mottelson \cite{BM75}.
Hereafter, we denote the textbook \cite{BM75} by ``BM2''.

The SBS is strictly defined only in infinite systems.
Therefore, in the beginning of this article
(sections~\ref{sec:unified_model}, \ref{sec:finite_size_effect},
and \ref{sec:collective_motion}),
we address the following basic questions.
\begin{enumerate}
\item What is the origin of nuclear deformation?
\item What is the meaning of the SBS in finite systems?
\item What kind of collective motion will emerge due to the SBS?
\end{enumerate}
These are important issues to understand the essence of
the nuclear structure physics.
We hope that these sections are useful, especially 
for students and non-practitioners.

The rotational motion is a collective motion emerged from the SBS,
corresponding to the massless Anderson-Nambu-Goldstone (ANG) mode
\cite{And52,And58-1,And58-2,Nam60,NJL61,Gol61}
in the infinite system.
It is approximately decoupled from the intrinsic motions, however,
the decoupling is never exact.
Moreover, as we mentioned in the beginning, we may generate
a nucleus spinning extremely fast in experiments.
Coupling between the rotational and intrinsic motions
produces a variety of phenomena.
We will discuss several related topics in
sections~\ref{sec:coupling}, \ref{sec:SD}, and \ref{sec:wobbling}.

The coupling introduces the Coriolis mixing among different bands.
The angular momentum dependence ($I$-dependence) of
the transition matrix elements is very sensitive to this, even at
low spin.
The unified model predicts a form of the $I$-dependent
intensity relation,
however, a systematic way of calculating intrinsic moments
entering in the intensity relation was missing.
We present a feasible microscopic method for the calculation of
the intrinsic moments using the cranking model at
an infinitesimal rotational frequency (section~\ref{sec:coupling}).

Low-lying vibrational modes of excitation strongly reflect the
underlying shell structure.
Therefore, the new shell structure produced by rapid rotation
and large deformation may significantly change their properties,
and could lead to new soft modes and instability.
Octupole vibrations at large angular momentum and in superdeformed bands
are discussed.
We present a possible banana-shaped superdeformation as a
consequence of the superdeformed shell structure (section~\ref{sec:SD}).

Nuclear wobbling motion, predicted by Bohr and Mottelson
(section 4-5 in BM2),
has been observed in $^{163}$Lu and neighboring nuclei.
This mode corresponds to a non-uniform three-dimensional rotation
and provides a clear signature of nuclear deformation without the
axial symmetry.
Microscopic analysis reveals an important role played by
the quasiparticle alignment.
In fact, without the alignment,
the wobbling motion cannot exist in $^{163}$Lu.
These issues and precession motion of the high-$K$ isomers
are discussed in section~\ref{sec:wobbling}.

For most of these studies, we use 
the quasiparticle-random-phase
approximation (QRPA) in the rotating shell model,
which we have developed for studies of rapidly rotating nuclei
\cite{SM82,MSM90,NMMS96,SS09}.
Further inclusion of the quasiparticle-vibration coupling
has been carried out for odd nuclei \cite{MSM88,NMM93,Mat14}.
The method is still very useful and illuminating to obtain
insights into nuclear structure in extreme conditions.
In the present article, we do not present details of the
theoretical models.
Instead, we would like to concentrate our discussion on
basic concepts and emergent phenomena.

\section{Unified model and spontaneous breaking of symmetry}
\label{sec:unified_model}

The atom is a finite-size quantum system,
composed of electrons bound by the Coulombic attraction of
the central nucleus.
The nucleus is also a quantum finite-size fermionic system
composed of nucleons.
In both systems, the independent-particle (single-particle) motion
is a prominent feature which leads to the ``shell model''.
In the first order approximation,
the constituent particles (electrons in the atom and nucleons in the nucleus)
freely move in the confining potential.
However, there is an obvious but important difference
between the atom and the nucleus.
Namely, the nuclear potential binding the nucleons is generated
by the nucleons themselves.

The electrons in the atom are bound by
the attractive Coulomb potential generated by the nucleus.
This potential is spherical, $-Ze^2/r$, in the atomic scale ($r\sim$\AA).
Although the repulsive interaction creates correlations
among the electrons, the strong attractive potential always produces a
restoring force which favors the spherical shape.
In contrast, the shape of the nuclear potential is determined by
the shape of the nucleus itself.
It is often referred to as {\it ``nuclear self-consistency''}.
Therefore, we expect that the nucleus may change its shape,
much easier than the atomic case.
In other words, the nucleus is rather ``soft'' and produces low-energy
``slow'' shape vibrations.

\subsection{Unified model Hamiltonian}

Bohr and Mottelson treated these shape degrees of freedom as collective 
variables $\alpha$ in addition to the single-particle
degrees of freedom $\xi$.
In general, the shape dynamics described by $\alpha$ is considerably
slower than the single-particle motion.
Thus, we could adopt a picture that the nucleons move in a one-body
potential $V(\xi,\alpha)$ which is specified by the nuclear shape $\alpha$.
The idea ends up with the unified-model Hamiltonian,
\begin{equation}
H=H_{\rm coll}(\alpha) + H_{\rm sp}(\xi) + H_{\rm int}(\xi,\alpha) ,
\label{unified_model}
\end{equation}
where $H_{\rm coll}$ is the {\it collective} Hamiltonian to describe
the low-energy shape vibrations.
$H_{\rm sp}$ corresponds to the {\it single-particle} (shell model) Hamiltonian
at the spherical shape, $H_{\rm sp}(\xi)=T_{\rm kin}(\xi) + V(\xi,\alpha=0)$.
$T_{\rm kin}$ is the {\rm kinetic} energy term and
the nuclear self-consistency requires
the potential $V(\xi,\alpha)$ to vary with respect to the shape $\alpha$.
The {\it interaction} between the collective and single-particle motions,
given by the third term $H_{\rm int}(\xi,\alpha)$,
is indispensable to take into account this important property of
nuclear potential.

\subsection{Symmetry breaking mechanism}

The coupling term in equation~(\ref{unified_model}) could lead the nucleus
to deformation.
This is associated with the SBS mechanism.
To elucidate the idea,
let us adopt a simple adiabatic (Born-Oppenheimer) approximation.
First, we solve the eigenvalue problem for
the Schr\"odinger equation for the variables $\xi$
with a fixed value of $\alpha$,
\begin{equation}
H_{\rm def}(\alpha) \ket{\phi_n(\alpha)}
 = \epsilon_n(\alpha) \ket{\phi_n(\alpha)} ,
\end{equation}
where $H_{\rm def}(\alpha)\equiv H_{\rm sp} + H_{\rm int}(\alpha)$.
This gives the adiabatic collective Hamiltonian,
$H_{\rm ad}^{(n)}(\alpha)=H_{\rm coll}(\alpha)+\epsilon_n(\alpha)$,
for each intrinsic eigenstate
$\phi_n(\xi;\alpha)\equiv\inproduct{\xi}{\phi_n(\alpha)}$.
$H_{\rm ad}^{(n)}$ is an effective Hamiltonian for
the collective variables $\alpha$.
The total wave function is given by a product of the intrinsic and
the collective parts \cite{Vil84},
$\Psi_n(\alpha,\xi)=\psi^{(n)}(\alpha)\phi_n(\xi;\alpha)$.

There are two possible mechanisms of the SBS in the unified model
to realize the deformed ground state with $\alpha\neq 0$.
When $\epsilon_0(\alpha)$ strongly favors the deformation,
even if $H_{\rm coll}(\alpha)$ has the potential minimum at $\alpha=0$,
the adiabatic potential in $H_{\rm ad}^{(0)}(\alpha)$
may have a deformed minimum.
Apparently, this mechanism requires 
the deformation-driving nature of $H_{\rm int}(\xi,\alpha)$,
which we call ``{\it coupling-driven mechanism}''.
On the other hand, there is another mechanism
which can deform the nucleus even if
the spherical shape ($\alpha=0$) is favored by the adiabatic
ground state $\epsilon_0(\alpha)$.
This is due to the additional coupling caused by
the kinetic term of $H_{\rm coll}$;
$T_{\rm kin}(\alpha)=-(1/2)B\partial_\alpha^2$.
We adopt units of $\hbar=1$ throughout the present article.
Roughly speaking,
the SBS takes place when the level spacing in $\epsilon_n(\alpha)$
is smaller than the additional coupling.
It is analogous to the Jahn-Teller effect in the molecular physics
\cite{JT37},
which we call ``{\it degeneracy-driven mechanism}''.

\subsection{Degeneracy-driven SBS; diagonal approximation}

In order to understand the degeneracy-driven mechanism, we
make the argument simpler, 
neglecting the off-diagonal elements,
$\bra{\phi_n(\alpha)}T_{\rm kin}(\alpha)\ket{\phi_0(\alpha)}$ ($n\neq 0$).
Integrating the intrinsic (single-particle) degrees $\xi$,
the effective Hamiltonian for the collective variable $\alpha$ is obtained as
\begin{equation}
H_{\rm eff}^{(0)}(\alpha)=\bra{\phi_0(\alpha)} H \ket{\phi_0(\alpha)}
=H_{\rm coll}^{(0)}(\alpha)+\epsilon_0(\alpha)+\Phi_0(\alpha),
\end{equation}
where $H_{\rm coll}^{(0)}$ is identical to $H_{\rm coll}$ except
that its kinetic energy is modified into $T_{\rm kin}^{(0)}(\alpha)= -(1/2)B
(\partial_\alpha + \inproduct{\phi_0}{\partial_\alpha\phi_0})^2$.
This is equivalent to introduction of a ``vector'' potential \cite{Ber84},
$A(\alpha)\equiv i\inproduct{\phi_0}{\partial_\alpha\phi_0}$.
If the coordinate $\alpha$ is one-dimensional, the ``vector'' potential
$A(\alpha)$ can be eliminated by a gauge transformation,
$\exp(i\int A(\alpha) d\alpha)$.
However, the following ``scalar'' potential remains.
\begin{eqnarray}
\Phi_0(\alpha)
&=&\frac{1}{2}B\bra{\partial_\alpha\phi_0}
 (1-\ket{\phi_0}\bra{\phi_0})
\ket{\partial_\alpha\phi_0} 
=\frac{1}{2}B\sum_{n\neq 0}
 \inproduct{\partial_\alpha\phi_0}{\phi_n}
 \inproduct{\phi_n}{\partial_\alpha\phi_0} 
\nonumber \\
&=&\frac{1}{2}B\sum_{n\neq 0}
\left|
\frac{\bra{\phi_n(\alpha)}(\partial_\alpha H_{\rm def}(\alpha))
\ket{\phi_0(\alpha)}}
     {\epsilon_0(\alpha)-\epsilon_n(\alpha)}
\right|^2 .
\label{Phi_0}
\end{eqnarray}
From equation~(\ref{Phi_0}), it is apparent that $\Phi_0(\alpha)$ is positive
and becomes large where the adiabatic ground state is approximately
degenerate in energy, $\epsilon_0\approx \epsilon_n$ ($n\neq 0$).
When the spherical ground state ($\alpha=0$) shows degeneracy,
it could be significantly unfavored by $\Phi_0(\alpha)$.
The system tends to avoid the degenerate ground state, which leads to
the SBS with nuclear deformation.

We would like to emphasize again that
the coupling between the collective (shape)
degrees of freedom $\alpha$ and the intrinsic (single-particle) motion
$\xi$ is essential to produce the nuclear deformation.
This is apparent for the coupling-driven mechanism, and is also true
for the degeneracy-driven case.
If the coupling term $H_{\rm int}(\xi,\alpha)$ is absent,
the adiabatic states $\phi_n(\xi)$ are independent of $\alpha$,
thus, produce no gauge potentials, $A(\alpha)=\Phi_0(\alpha)=0$.
We also note here that the present argument on the degeneracy-driven
(Jahn-Teller) mechanism explains why the instability of a spherical state
occurs, but not what kind of deformation takes place.
This will be discussed in sections~\ref{sec:shell_structure}
and \ref{sec:banana_SD}.

\subsection{Field coupling}

The oscillation of the variable $\alpha$ correspond to the shape vibration.
Thus, it can be quantized to a boson operator.
In order to describe the vibrational motion associated with $\alpha$,
we introduce a boson space with the $n$-phonon state $\ket{n}$.
When $\alpha$ is small,
we may linearize the coupling term in equation~(\ref{unified_model})
with respect to $\alpha$ as
\begin{equation}
H_{\rm int}(\xi,\alpha)=-\kappa\alpha F(\xi) ,
\label{H_int}
\end{equation}
where $\kappa$ is a coupling constant which depends on the
normalization of $\alpha$ and $F$.
If the operator $F$ is given, the normalization of $\alpha$ is
usually chosen as follows.
The action of the one-body operator $F$ on the ground state
(a Slater determinant) produces many one-particle-one-hole states;
$F\ket{\Phi_0}=\sum_{ph} \ket{\Phi_{ph}}\bra{\Phi_{ph}}F\ket{\Phi_0}$.
This is identified with the operation of $\alpha$
in the collective (boson) space:
\begin{equation}
\sum_{ph}|\bra{\Phi_{ph}}F\ket{\Phi_0}|^2 = |\bra{n=1}\alpha\ket{n=0}|^2 .
\label{normalization}
\end{equation}
The coupling constant $\kappa$ can be also determined by this
self-consistency.
See chapter 6 of BM2 for details of the field coupling techniques.

If the matrix elements of $F$ are identical to
those of $\alpha$ as in equation~(\ref{normalization}),
the field coupling~(\ref{H_int}) can be interpreted as
an residual two-body interaction
\begin{equation}
H_{\rm res}(\xi)=-\frac{1}{2}\kappa F^2 .
\label{separable_interaction}
\end{equation}
This kind of separable effective interactions have been extensively
adopted in nuclear structure studies.
Among them, the most famous one is the pairing-plus-quadrupole model
\cite{BK65,KB68-1,BS69},
which was originally proposed by Bohr, Mottelson, and their colleague.
It represents two kinds of important low-energy correlations in nuclei;
One is the quadrupole correlations, $F\sim r^2Y_{2\mu}$,
which are inspired by existence of low-lying $2^+$ vibrational
excitations in even-even nuclei.
Another correlation is the pairing,
$F\sim P+P^\dagger$ and $F\sim P-P^\dagger$
where $P$ is the pair annihilation operator.
This is important in heavy nuclei in which the nuclear superfluidity 
associated with the pair condensation is well established \cite{BB05}.
We also adopt this separable form as a residual interaction for
the QRPA calculations in sections~\ref{sec:coupling}, \ref{sec:SD},
and \ref{sec:wobbling}.

\section{Finite-size effect}
\label{sec:finite_size_effect}

Nuclei on earth are of finite size ($R<10$ fm)
with finite number of nucleons ($A<300$).
Strictly speaking,
the SBS in the ground state is realized in the infinite system.
For finite systems, the quantum fluctuation associated with the zero-point
motion restores the broken symmetry.
Thus, the symmetry-broken state is not stable for finite systems,
in a rigorous sense.
However, the SBS is ubiquitous in macroscopic objects in nature,
which are made of big but {\it finite} number of particles.
Thus, everybody agrees that
the zero-point motion to restore the symmetry
can be safely neglected in
the macroscopic number, say $A\sim 10^{23}$.
Then, how about the case of $A\sim 200$?

\subsection{Finite correlation time}
\label{sec:finite_correlation_time}

Let us consider a deformed nucleus and the single-particle states
$\phi_i^0$ in the deformed Nilsson potential.
The deformed ground state is simply assumed to be a Slater determinant,
$\ket{\Phi_0}\equiv\det\{\phi_1^0\phi_2^0\cdots\phi_A^0\}$.
If we rotate the nucleus by angle $\theta$, we have a state
$\ket{\Phi_\theta}=
\det\{\phi_1^\theta\phi_2^\theta\cdots\phi_A^\theta\}$.
where $\phi_i^\theta=\hat{R}(\theta)\phi_i^0$ with the rotation operator
$\hat{R}(\theta)$.
Each single-particle state $\phi_i^\theta$ in the tilted Nilsson potential
can be expanded in terms of the untilted state $\phi_i^0$, as
$\phi_i^\theta=\sum_j^\infty c_{ij}^\theta \phi_j^0$.
When the angle $\theta$ is small,
we can estimate the diagonal coefficients as $|c_{ii}|\sim 1-c|\theta|$
with $c>0$, and the off-diagonal ones ($i\neq j$) as $|c_{ij}|\sim O(\theta)$.
As far as the nucleon number $A$ is finite,
the tilted ground state $\ket{\Phi_\theta}$ can be written in terms of
the untilted Nilsson basis, $\{ \phi_i^0 \}$.
This is due to the fact that $\ket{\Phi_0}$ and $\ket{\Phi_\theta}$
belong to the same Hilbert space.

However, in the limit of $A\rightarrow\infty$,
this is no longer true.
The tilted ground state is expanded in terms of
the untilted Slater determinants as
\begin{equation}
\ket{\Phi_\theta}=\det\{\phi_1^\theta\cdots\phi_A^\theta\}
=\sum_{j_1,\cdots, j_A} C_{j_1\cdots j_A}
\det\{\phi_{j_1}^0\cdots\phi_{j_A}^0 \} ,
\end{equation}
where $C_{j_1 j_2\cdots j_A}=c_{1j_1}c_{2j_2}\cdots c_{Aj_A}$.
For a small value of $\theta$,
the largest coefficient among $\{ |C_{j_1\cdots j_A}|\}$
is apparently $|C_{1\cdots A}|$ whose absolute magnitude is
$|C_{1\cdots A}|\sim (1-c|\theta|)^A$.
Therefore, all the coefficients $C_{j_1\cdots j_A}$
vanish exponentially as functions of $A$.
This means that $\ket{\Phi_0}$ and $\ket{\Phi_\theta}$ belong to
different Hilbert spaces at $A\rightarrow\infty$,
thus, $\ket{\Phi_\theta}$ is no longer expandable
in terms of the untilted Slater determinants.
In other words, the deformed {\it infinite} nucleus never
rotates.

The same issue can be examined in terms of the excitation spectra.
The rotational spectra of deformed nuclei show $E_I=I(I+1)/2{\cal J}$,
in which the moment of inertia $\mathcal{J}$
is approximately order of $A^{5/3}$.
The rotational motion is {\it quantized} due to the finiteness of ${\cal J}$.
In the limit of $A\rightarrow\infty$,
the excitation spectra becomes {\it gapless} and 
the ground state ($I=0$) is degenerate with other states ($I\neq 0$).
Therefore, an infinitesimally weak external field can fix its orientation
by superposing states with different $I$.

Now, let us come back to the question,
"how about heavy deformed nuclei?".
As far as $A$ is finite, the ``tilted'' and ``untilted''
Hilbert spaces are equivalent.
The zero-point fluctuation may connect
$\ket{\Phi_0}$ and $\ket{\Phi_\theta}$,
thus, the wave packet $\ket{\Phi_0}$
loses its direction in finite correlation time.
If this time scale is significantly larger than that of the single-particle
motion $\tau_F\sim 10^{-22}$ s, 
we can claim that the SBS takes place and the nucleus is deformed.
In fact, this condition is well satisfied for heavy nuclei.
Let us limit the orientation of the deformed nucleus to
an angle range of unity ($\Delta\theta\sim 1$),
then, the quantum fluctuation produces the angular momentum with the
magnitude of $\Delta I \sim (\Delta\theta)^{-1} \sim 1$.
This leads to the correlation time,
$\tau_{\rm cor}\sim {\cal J}/\Delta I \sim {\cal J}$,
that amounts to $10^{-19}$ s for typical deformed actinide nuclei.
This argument is consistent with the vanishing behavior of the
coefficients $C_{j_1\cdots j_A}$.
Suppose the overlaps $|\inproduct{\phi_i^0}{\phi_i^\theta}|\sim 0.9$ for
$i=1,\cdots,A$,
then, we have $C_{12\cdots A}\sim (0.9)^A \sim 7\times 10^{-10}$
for $A=200$.
Therefore, the rotational fluctuation is significantly hindered for
heavy deformed nuclei.
These simple exercises also tell us that,
the concept of SBS has a greater significance for
nuclei with larger $A$ and larger deformation.

From a similar argument
replacing the single-particle states $\phi_i^0$ in the Slater determinant
$\det\{\phi_{j_1}^0\cdots\phi_{j_A}^0 \}$ by those in a spherical potential,
we may understand why the description based on the ``symmetry-broken''
deformed basis is important.
It is apparent that,
if we adopt a spherical shell model basis for such heavy well-deformed nuclei,
we need to treat very small coefficients, $C_{j_1\cdots j_A}$,
with enormous number of basis states.
In the limit of $A\rightarrow\infty$, this treatment becomes impossible.
The impossibility here is in a strict sense,
not in a practical sense due to computational limitation.
Thus, instead of superposing the ``symmetry-preserving'' (spherical)
Slater determinants, the theories of restoring broken symmetry,
such as the projection method,
have been extensively developed in nuclear physics,
to take into account effects of the zero-point fluctuation \cite{RS80}.
The usefulness of these symmetry restoration approaches has been
recognized recently in other fields \cite{YL07}.

\subsection{Zero-point motion and shell effect}

As we have mentioned in section~\ref{sec:finite_correlation_time},
the finiteness leads to
the finite correlation time and the finite energy gap in the excitation 
spectra.
In the symmetry restoration mechanism, the zero-point fluctuation
associated with the ANG mode is a key element.
In this subsection, we discuss effects of other zero-point motions in
finite systems, which could hinder the SBS.
The zero-point kinetic energy of nucleons is roughly given as
$T_\textrm{\sc zpe}/A \sim 1/(mR^2) \sim 10$ MeV.
This is comparable to the nuclear binding energy $B/A\sim 8$ MeV
and has a non-negligible effect.
In fact, since the nucleons are fermions, the Fermi energy is even larger,
$t_F\sim k_F^2/(2m)\sim 40$ MeV.
The zero-point (Fermi motion) kinetic energy
generally favors the ``symmetry-preserving'' state with
a uniform and spherical density distribution.
Since this competes with the SBS driving effect,
the SBS which occurs in the thermodynamical limit may not occur
in finite systems.
The interplay between the zero-point motion and the interaction
leads to interesting phenomena in nuclei.

The shell effect is a kind of finite-size effect in many fermion systems
and is an indispensable factor in the low-energy
nuclear structure.
The prominent deformation hindrance effect can be found
at the spherical magic numbers.
The ground states of those magic nuclei favor spherical shape.
Nevertheless,
most of the spherical nuclei show the shape coexistence phenomena.
For instance, the even-even spherical nuclei often have deformed excited
$0^+$ states at very low energy.
It is prevalent in many semi-magic nuclei, and even true for
some doubly magic nuclei.
In contrast, as far as we know, when the ground state is deformed,
excited $0^+$ spherical states have not been clearly identified.
In the Strutinsky shell correction method \cite{Bra72,RS80},
the shell effect is regarded as an origin of the nuclear deformation.
It might be proper to say this in an opposite way;
the heavy nuclei are ``genetically'' deformed,
and some special nuclei become spherical
because of the finite-size (spherical shell) effect to hinder the SBS.

\subsection{Shell structure and soft modes}
\label{sec:shell_structure}

When the symmetry breaking takes place and the nucleus is deformed,
what kind of shape is realized?
This depends on the underlying shell structure.
Let us present a simple argument  based on the one given by
by Bohr and Mottelson (pp. 578$-$591 in BM2).
For a spin-independent spherical potential,
the single-particle energy is characterized
by the radial quantum number $n$ and the orbital angular momentum $l$,
$\epsilon(n,l)$.
When we change $n$ and $l$ from a certain value $(n_0,l_0)$.
\begin{equation}
\epsilon(n,l)
=
\epsilon(n_0,l_0)+ \Delta n
\left(\frac{\partial\epsilon}{\partial n}\right)_0
+\Delta l
\left(\frac{\partial\epsilon}{\partial l}\right)_0
+ \cdots ,
\label{epsilon}
\end{equation}
where $\Delta n=n-n_0$ and $\Delta l=l-l_0$.
Since $n$ and $l$ take only integer numbers,
the ratio,
$a:b\equiv
({\partial\epsilon}/{\partial n})_0:({\partial\epsilon}/{\partial l})_0$
plays a very important role.
If the ratio $a:b$ is rational, we can choose
$a$ and $b$ as the integer numbers.
Then, in the linear order (\ref{epsilon}),
$\epsilon(n,l)$ and $\epsilon(n\pm mb, l\mp ma)$ are degenerate,
where $m$ is an integer number.

Now, let us define the shell frequency as
\begin{equation}
\omega_{\rm sh}\equiv 
\frac{1}{a}\left(\frac{\partial\epsilon}{\partial n}\right)_0
=\frac{1}{b}
\left(\frac{\partial\epsilon}{\partial l}\right)_0 .
\end{equation}
There are degenerate single-particle energies at intervals of
$\omega_{\rm sh}$.
Larger integers $a$ and $b$ correspond to a smaller $\omega_{\rm sh}$.
Therefore, the prominent shell structure with a large shell gap
$\omega_{\rm sh}$ should be associated with the small integers $(a,b)$.
For instance, the isotropic harmonic oscillator potential
has the $a:b=2:1$ shell structure, with the constant $\omega_{\rm sh}$.
The Coulomb potential has the strict $a:b=1:1$, with the energy-dependent 
$\omega_{\rm sh}$.
In general, the degeneracy is approximate and the ratio $a:b$ may change
according to the location of the Fermi level.

The derivatives, $(\partial\epsilon/\partial n, \partial\epsilon/\partial l)$,
correspond to the (angular) frequencies in the classical mechanics;
$\partial\epsilon/\partial n$ is the frequency of the radial motion,
while
$\partial\epsilon/\partial l$ is that of the angular motion.
The integer ratio $(a,b)$ of the frequencies means that the classical orbit
is closed (periodic).
Therefore, the quantum shell structure is closely related to
the classical periodic orbits.

Since the nuclear potential somewhat resembles
the harmonic oscillator potential,
the shell structure associated with $a:b=2:1$
is prominent.
The $2:1$ periodic orbit in the harmonic oscillator potential
is the elliptical orbit.
When there are many valence nucleons in the degenerate levels,
the short-range attractive interaction favors their maximal overlap,
which eventually leads to the SBS
to an ellipsoidal (quadrupole) shape.
In the quantum mechanical terminology,
we may say that the coupling among the degenerate single-particle
levels with $\Delta l=2$ produces a soft mode.
If the number of valence nucleons becomes large,
this correlation may produce the quadrupole deformation.

The spin-orbit potential decreases
the frequency $\partial\epsilon/\partial l$
for the single-particle levels of $j=l+1/2$.
This could lead to a new shell structure of $a:b=3:1$
among the levels of the $j=l+1/2$.
The $3:1$ frequency ratio corresponds to classical periodic orbits of
the triangular shape,
since the radial motion oscillates three times during the single
circular motion.
Thus, for heavy nuclei in which the high-$j$ single-particle levels
($j=l+1/2$) are located near the Fermi level,
the approximate degeneracy of the $\Delta l=3$ levels 
may result in the octupole instability
in open-shell configurations.
For example,
the neutron-deficient actinide nuclei show typical spectra of
the alternating parity band,
which are understood as a realization of the pear-shaped deformation
of $Y_{30}$ type \cite{BN96}.

The investigation of the classical periodic orbits is 
useful to identify a soft mode and a favorable shape.
The SBS toward the quadrupole deformation in open-shell nuclei
is nicely explained
in this simple argument.
However, it is more difficult to explain the fact that most nuclei have
the prolate shape, not the oblate shape.
There have been a number of works on this issue \cite{Fri90,TS01,HM09,TOST11}.
According to the classical periodic orbits, a recent analysis
sheds new light on the prolate dominance in nuclei \cite{Ari12}.

\subsection{Fermi motion and nuclear self-consistency}
\label{sec:self-consistency}

In nuclei, the kinetic energy of nucleons' Fermi motion is very large.
Adopting the harmonic oscillator potential model,
a simple estimate of the total kinetic energy is given by
\begin{equation}
T_0=\frac{1}{2}E_{\rm HO}
=\frac{1}{2}\sum_{k=x,y,z}\omega_k\Sigma_k,
\quad\quad
\Sigma_k\equiv\sum_{i=1}^A (n_k+\frac{1}{2})_i ,
\end{equation}
where $n_k$ ($k=x,y,z$) are the oscillator quantum numbers of
the single-particle states.
For a spherical nucleus ($\omega_x=\omega_y=\omega_z=\omega_0$)
filling the levels up to $n_x+n_y+n_z=N_{\rm max}$,
this amounts to 
\begin{equation}
T_0=(1/4)\omega_0
(N_{\rm max}+1)(N_{\rm max}+2)^2(N_{\rm max}+3) .
\end{equation}
Taking $Z=N=40$ ($^{80}$Zr, $N_{\rm max}=3$), this gives
$T_0=150\omega_0\sim 1.43$ GeV,
with a standard value of $\omega_0\approx 41$A$^{-1/3}\sim 9.5$ MeV.
If we deform the harmonic oscillator to a prolate/oblate shape with
$(\omega_x,\omega_y,\omega_z)=(e^\eta,e^\eta,e^{-2\eta})\omega_0$,
the kinetic energy becomes
\begin{equation}
T(\alpha)=\frac{2e^\eta+e^{-2\eta}}{3}T_0 ,
\label{T_alpha}
\end{equation}
which has the minimum value at the spherical shape $\eta=0$.
According to equation~(\ref{T_alpha}),
a moderate prolate deformation of $e^\eta=1.1$ will
produce the increase in the kinetic energy by about 1 \%.
However, since $T_0$ is very large, this 1 \%\ increase is
significant, such as 14 MeV for $Z=N=40$.
However, the deformed ground state in $^{80}$Zr
is suggested by experiments observing a ground-state rotational band
\cite{Lis87}.
How does the nucleus compensate this large increase in kinetic energy?

The solution to this problem
is again attributed to the nuclear self-consistency.
In the harmonic oscillator model, the self-consistency condition, that
the deformation of the potential is equal to that of the density
distribution,
can be simply expressed by equation (4-115) in BM2,
\begin{equation}
\omega_x\Sigma_x =
\omega_y\Sigma_y =
\omega_z\Sigma_z .
\label{self_consistency_ho}
\end{equation}
Namely, when the nuclear potential is deformed as
$(\omega_x,\omega_y,\omega_z)=(e^\eta,e^\eta,e^{-2\eta})\omega_0$,
the configuration of the ground state should change accordingly,
$(\Sigma_x,\Sigma_y,\Sigma_z)=(e^{-\eta},e^{-\eta},e^{2\eta})\Sigma_0$.
Since the momentum distribution in the harmonic oscillator potential
model can be calculated as
$\langle p_k^2 \rangle = m  \omega_k \Sigma_k$ ($k=x,y,z$),
the self-consistency condition~(\ref{self_consistency_ho})
means the isotropic momentum (velocity) distribution (no deformation in the
Fermi sphere).
In other words, the shape of the nucleus is specified by
the minimization of the kinetic energy which is equal to the
isotropic velocity distribution.

This indicates the importance of configuration rearrangements
in low-energy collective dynamics.
When the nuclear deformation is changed as
$(\omega_x,\omega_y,\omega_z)\rightarrow
(e^\eta\omega_x,e^{\eta'}\omega_y,e^{-\eta-\eta'}\omega_z)$,
the configuration should follow as
$(e^{-\eta}\Sigma_x,e^{-\eta'}\Sigma_y,e^{\eta+\eta'}\Sigma_z)$,
to keep the Fermi sphere spherical.
In order to change the configuration, we need two-particle-two-hole
excitations, to annihilate
a time-reversal pair of nucleons in a certain single-particle orbit
and create a pair in another orbit.
Therefore, we expect that,
during the shape evolution at low energy,
the pairing interaction plays a dominant
role in dynamical change of the configuration.
This was supported by experimental data that the spontaneous fission
life times of even-even nuclei are much shorter than those of
odd and odd-odd nuclei \cite{Swi82}.

According to the nuclear self-consistency,
each configuration has its optimal shape.
We may think about possibilities of realizing
different shapes corresponding to different configurations
in the same nucleus.
This phenomenon is called ``shape coexistence''.
For instance, in the harmonic oscillator model of $^{80}$Zr with $N=Z=40$,
in addition to the spherical configuration
($\Sigma_x=\Sigma_y=\Sigma_z$, $\omega_x=\omega_y=\omega_z$),
the self-consistency condition~(\ref{self_consistency_ho})
is also satisfied with the superdeformed configuration
($2\Sigma_x=2\Sigma_y=\Sigma_z$, $\omega_x=\omega_y=2\omega_z$).
In fact, the shape coexistence phenomena have been observed in
many areas throughout the nuclear chart \cite{HW11}.

\subsection{Fermi sphere in the rotating frame}

This idea of the isotropic velocity distribution
can be extended into the one 
in the rotating frame (p.79 in BM2).
The local velocity in the rotating frame,
\begin{equation}
\vec{v}\equiv \vec{p}/m - (\vec{\omega}_{\rm rot}\times\vec{r}),
\label{v_rotating_frame}
\end{equation}
has an isotropic distribution $\rho_{\vec{r}}(v)$ at each $\vec{r}$.
The isotropic velocity distribution means no net current relative to
the rotating frame, which ends up with a rigid-body value for
the moment of inertia.
The deformed nucleus
would have a rigid-body value of moment of inertia if the pairing
correlations were absent.

The transformation from the laboratory frame to the rotating frame
leads to the cranking Hamiltonian
\begin{equation}
H'=H-\vec{\omega}_{\rm rot} \cdot \vec{J} ,
\label{cranking_hamiltonian}
\end{equation}
where $\vec{J}$ is the total angular momentum.
The velocity-dependent terms (kinetic energy and the centrifugal potential)
in the rotating frame can be written as
$p^2/(2m) -\vec{\omega}_{\rm rot}\cdot(\vec{r}\times\vec{p})
 = mv^2/2 - m(\vec{\omega}_{\rm rot}\times\vec{r})^2/2$
where $\vec{v}$ is given by equation~(\ref{v_rotating_frame}).
This confirms the $\vec{v}$-dependence in $H'$ is isotropic ($\propto v^2$).
This isotropic velocity distribution is still valid in
rotating nuclei \cite{SM84}.
The cranking model~(\ref{cranking_hamiltonian}) plays
a key role in physics of high-spin nuclear structure
(sections~\ref{sec:coupling}, \ref{sec:SD}, and \ref{sec:wobbling}).

\section{SBS and collective motions}
\label{sec:collective_motion}

A broken continuous symmetry leads to the emergence of two types of
collective excitations.
One is the massless ANG mode
and the other is the massive Higgs mode \cite{PV15}.
Therefore, properties of the collective motions significantly change
before and after the SBS takes place.
In nuclei, we can observe them in the excitation spectra.
We can even see how the ANG and Higgs modes appear and evolve
from soft modes.

\subsection{Rotational motion; ANG mode}

The ANG mode is a gapless (massless) mode in the infinite system.
For the case of nuclear deformation,
the ANG mode correspond to the rotational motion of the deformed nucleus.
Because of the finiteness,
the spectrum is not exactly gapless, however,
shows a gradual emergence of the ``quasi-degenerate'' rotational spectra.

In Fig. 6-31 of BM2,
a typical example for even-even Sm isotopes ($^{144-154}$Sm)
is presented.
The $^{144}$Sm nucleus has the magic neutron number $N=82$.
Its ground state ($0^+$) is spherical and
the first excited state is located at excitation energy of 1.63 MeV.
Keeping the proton number the same and increasing the neutron number
two by two, we clearly observe the following:
\begin{enumerate}
\item The first $2^+$, $4^+$, $\cdots$ ($2I$) states lower their
 excitation energies.
 Eventually, a rotational band is formed to present the
 excitation spectra, $E_I\propto I(I+1)$.
\item The second $0^+$ and the second $2^+$ states
 lower their energies in the beginning.
 However, they stop decreasing at $N=88$ ($^{150}$Sm).
\item Additional rotational bands are formed on top of
 the second $0^+$ and $2^+$ states for $N\geq 90$ ($^{152,154}$Sm).
\end{enumerate}
In $^{154}$Sm, the excitation energy of the first $2^+$ state
is only 82 keV.
This is 1/20 of that in $^{144}$Sm and we may say that it is
approximately degenerate in the ground state.
Moreover, there appear five members ($0^+,\cdots,8^+$)
of rotational bands below 1 MeV of excitation.
It should be noted that similar phenomena
are observed in many region of nuclear chart,
when the neutron (proton) numbers are going away from the
spherical magic number.

A regular pattern of rotational spectra allows us to distinguish
the intrinsic excitations and the rotational motion.
A rotational band is constructed based on each intrinsic excitation
from the ground state.
From these observation, we may think of the Hamiltonian subtracting
the rotational energy, $H'\equiv H-{\vec{J}^2}/(2{\cal J})$.
$H'$ conserves the rotational symmetry,
however, the member of the rotational bands ($0^+$, $2^+,\cdots$)
will be degenerate in energy.
Then, a deformed wave-packet state which violates
the rotational symmetry becomes an eigenstate of the Hamiltonian $H'$.

The number of activated rotational degrees of freedom depends
on the nuclear shape.
For axially symmetric spheroidal shape, there is no collective
rotation around the symmetry axis.
In other words, the angular momentum component along the symmetry
axis (called $K$ quantum number in the following)
is purely determined by the intrinsic motion.
In this case, the two rotational axes perpendicular to the symmetry
($z$) axis are possible,
but they are equivalent in the sense that
they have equal moment of inertia, ${\cal J}_x={\cal J}_y$.
In contrast,
if the nucleus has an equilibrium shape away from the axial symmetry
(triaxial shape),
the collective rotations about three axes are all activated,
and they can have different moments of inertia,
${\cal J}_x$, ${\cal J}_y$, and ${\cal J}_z$.
We may expect that the rotational spectra become richer and more complex.
The wobbling motion is known to be a typical mode of excitation
in the triaxial nuclei,
that will be discussed in section~\ref{sec:wobbling}.

\subsection{Beta and gamma vibrations; amplitude (Higgs) mode}
\label{sec:beta_gamma_vib}

The quadrupole ($\lambda=2$) vibrations
produce $2^+$ excitations in spherical nuclei.
When the SBS takes place to produce
the prolate (spheroidal) ground state,
among five $\alpha_{2\mu}$ ($\mu=-2,\cdots,2$),
the two shape degrees ($\beta$ and $\gamma$) remain,
and rest of the degrees of freedom are absorbed in the rotational motion
(Euler angle $\Theta$).
For an axially symmetric ground state,
the normal modes can be classified by the vibrational angular
momentum along the symmetry axis, which is often denoted by
the quantum number $K$.
The $\beta$ and $\gamma$ vibrations correspond to $K^\pi=0^+$ and $2^+$,
respectively.
Note that the $K^\pi=1^+$ low-lying vibration does not exist, because
it corresponds to the rotation of the whole nucleus.

The $\beta$ vibration around the SBS minimum is associated with
a collective amplitude mode of order parameter with a finite energy gap,
in contrast to the ``gapless'' rotational motion.
This type of excitation is often referred to as the
``amplitude (Higgs) mode''~\cite{PV15}.
A number of those candidates have been observed in well-known
deformed regions, such as the rare-earth and actinide regions.
For instance, in the rare-earth region, the excitation energies
of $\beta$ vibration candidates are found at $E_x\approx 1$ MeV.
However, their $B(E2)$ values from the ground states to the $2^+$ states
in the $\beta$-vibrational bands are not large in most cases,
typically a few Weisskopf units.
Instead, strong population by the pair transfer reaction has been
observed in many $\beta$-vibration candidates.
Therefore, their true nature is still mysterious
and currently under debate~\cite{Gar01}.
We should note that
an important role of the Coriolis coupling in the $\beta$ vibrations
has recently been pointed out \cite{MU16}.
See also figure \ref{fig:beta} in section \ref{sec:GIR}.

In contrast, the $\gamma$ vibrations,
whose excitation energies are also around 1 MeV,
show $B(E2)$ values significantly larger than
the Weisskopf units.
Thus, the collective nature of the $\gamma$ vibration
is well established.
Effects of their coupling to the rotational motion have been
also studied within the generalized intensity relation
(section~\ref{sec:GIR_applications}).

For nuclei with the prolate shape,
a naive geometric consideration may predict that the vibrational frequency
along the symmetry axis ($K=0$) is lower than that of the $K\neq 0$.
This is true for high-frequency giant quadrupole resonance \cite{HW01}.
However, the low-lying $\beta$ and $\gamma$ vibrations do not
follow this simple expectation (section 6-3b in BM2).
They are much more sensitive to underlying shell structure.

\subsection{Octupole vibrations; Negative-parity modes}
\label{sec:octupole_vib}

Octupole vibrations ($\lambda=3$)
with negative parity have been systematically
observed in spherical and deformed nuclei.
In spherical nuclei, it produces $3^-$ state.
The most typical example is perhaps that in $^{208}$Pb.
It is split into four different normal modes with $K^\pi=0^-\sim 3^-$
in deformed nuclei.
Again, the geometric expectation for their ordering is not applicable
to low-frequency octupole vibrations;
Namely, the $K^\pi=0^-$ vibrational state is not necessarily the lowest
among the multiplet.
The rotational band is formed on top of the bandhead with the spin
$I=K$ for $K=1\sim 3$ and $I=1$ for $K=0$.

\subsection{A microscopic tool;
quasiparticle-random-phase approximation (QRPA)}

In normal degenerate Fermi systems, the most basic mode of excitation
at low energy
corresponds to the one-particle-one-hole (1p1h) excitations.
When the pairing correlations produce the pair-condensed (BCS-like)
ground state, the 1p1h excitations should be replaced by
the two-quasiparticle (2qp) excitations.
The quasiparticle, which is mixture of particle and hole states,
is usually defined as an eigenstate of
the Hartree-Fock-Bogoliubov (HFB) equation \cite{RS80,BR86}.
The ground state corresponds to the quasiparticle vacuum state.
The 2qp excitations include not only 1p1h states, but also
two-particle and two-hole states
which correspond to states in neighboring nuclei with $A\pm 2$.
The odd-$A$ nuclei are expressed by one-quasiparticle states
based on the quasiparticle vacuum.

The collective excitations, such as $\beta$, $\gamma$, and
octupole vibrations in sections \ref{sec:beta_gamma_vib}
and \ref{sec:octupole_vib}, are approximately given by
superposition of many 2qp excitations.
The most successful theory for this purpose is the QRPA \cite{RS80,BR86},
which can describe both collective and non-collective modes
of excitation.
The QRPA contains backward amplitudes corresponding to
2qp annihilation on the correlated ground state,
and respects the symmetry of the Hamiltonian \cite{RS80,BR86}.
The limitation of QRPA is associated with its small-amplitude nature.

The HFB equations with the cranking Hamiltonian $H'$ 
(\ref{cranking_hamiltonian}) is often utilized for studies of
high-spin nuclear structure.
The QRPA calculation with the cranking Hamiltonian $H'$
is able to describe the rotational coupling effects,
such as the alignment and stretching, on
the collective and the non-collective excitations.
Some examples of the QRPA calculations with $H'$ are presented in
the following sections \ref{sec:coupling}, \ref{sec:SD},
and \ref{sec:wobbling}.

\section{Coriolis coupling to intrinsic motions}
\label{sec:coupling}

The SBS of the translational symmetry produces the ANG mode of
the center-of-mass motion.
Since it is exactly decoupled,
the intrinsic motions are not affected by the speed of
the nucleus in the accelerator.
On the other hand, the rotational motion is not exactly decoupled,
thus, the Coriolis and centrifugal effects influences
intrinsic structure.

In the unified model,
as is mentioned in section~\ref{sec:beta_gamma_vib},
the five quadrupole variables
$\alpha_{2\mu}$ ($\mu=-2,\cdots,2$) are represented by
$(\beta,\gamma)$ and three Euler angles $\Theta$.
Accordingly, the total wave function is given by a product
of the intrinsic, the vibrational, and the rotational parts.
If the shape fluctuation is neglected,
one can write the total wave function as a product of
the rotor and the intrinsic parts.
When the nucleus has the axially symmetric shape,
the intrinsic state has a good $K$-quantum number, $|K_n)$.
\begin{equation}
\ket{\Psi_{K_n IM}}=\ket{K_n IM}\otimes |K_n) ,
\label{unified_model_wf}
\end{equation}
where the rotor wave function is given by
\begin{equation}
\inproduct{\Theta}{KIM}=
\left(\frac{2I+1}{8\pi^2}\right)^{1/2}\mathcal{D}^I_{\lambda K}(\Theta) .
\end{equation}
The additional $\mathcal{R}$ invariance
requires the symmetrization of
equation~(\ref{unified_model_wf}) for $K_n\neq 0$;
$\{\ket{\Psi_{K_n IM}} + (-1)^{I+K} \ket{\Psi_{\bar{K}_n IM}}\}/\sqrt{2}$.
The quantum nature of the angular momentum is properly treated
in this rotor wave function.
The Coriolis coupling,
which mixes states with different $K$ quantum numbers,
can be treated in a perturbative manner.
Chapter 4 of BM2 presents extensive discussion on this subject.

On the other hand, in the high-spin limit $I\rightarrow\infty$, 
the semiclassical approximation works well.
The rotational frequency $\vec\omega_{\rm rot}$ is introduced,
which leads to the cranking model~(\ref{cranking_hamiltonian}).
Especially, when the direction of $\vec\omega_{\rm rot}$ is
parallel to a body-fixed principal axis $x$,
we have a uniform rotation $\omega_{\rm rot}(t)=\textrm{const}$.
\begin{equation}
H'=H-\omega_{\rm rot} J_x .
\label{cranking_hamiltonian_1D}
\end{equation}
This one-dimensional cranking model has been extensively applied
to high-spin nuclear structure problems with a tremendous success.
The non-linear effects of rotation are automatically 
taken into account in the intrinsic structure,
which reproduces a number of striking high-spin phenomena,
such as back-bending, alignment, and band termination.
A drawback is the missing quantum nature of rotation,
particularly important at low spin.

\subsection{Quantization of the cranking model at low spin}
\label{sec:GIR}

In the semiclassical approximation, the direction of
$\vec\omega_{\rm rot}\ (\vec{I})$
is assumed to be the $x$ axis of both the intrinsic
(body-fixed) and the laboratory (space-fixed) frames.
The multipole operator $\tilde{Q}_{\lambda\mu}$,
in which $\mu$ is defined with respect to the $x$ axis,
changes the angular momentum $I$ to $I+\mu$.
Thus, a transition matrix element
between states with the angular momenta $I_i$ and $I_f$ is simply given by
$\bra{f}\tilde{Q}_{\lambda \mu}\ket{i}$, where
$\mu$ should be equal to $\Delta I=I_f-I_i$.
This is a good approximation at the high-spin limit
($I,\omega_{\rm rot}\rightarrow\infty$).

In contrast, at the low-spin limit ($I\sim 0$),
the angular momentum is coupled to the deformation, 
thus, the $K$ quantum number along the symmetry ($z$)
axis is a good quantum number.
In this limit, the multipole operator ${Q}_{\lambda \nu}$,
defined with respect to the $z$ axis,
changes the $K$ quantum number by $\nu=\Delta K=K_f-K_i$.
In addition, the quantum mechanical nature of rotation is important
at low spin.
A perturbative expansion with respect to $I$ in the unified model
produces a specific $I$-dependence for the transition matrix element
({\it generalized intensity relations} in BM2).
To complete the intensity relation beyond the leading order,
we need to determine matrix elements of intrinsic operators
which take into account the Coriolis and centrifugal effects.
There is no systematic method to calculate these intrinsic matrix elements
in the unified model.

The cranking model~(\ref{cranking_hamiltonian_1D}), on the other hand,
is capable of microscopic treatment of
the rotational coupling to the intrinsic structure.
However, the semiclassical nature of the cranking model
forbids us to obtain the correct $I$-dependent intensity relations
at low spin.
This is mostly due to missing kinematics of the angular momentum algebra.
We present here a feasible prescription to recover the quantum mechanical
effect, that
enables us to calculate matrix elements of the intrinsic moments in
the generalized intensity relations.

\subsubsection{Generalized intensity relations}

The main idea is as follows \cite{SN96}.
In the high-spin limit, the cranking treatment becomes accurate and
the matrix elements of a multipole operator
$Q^{\rm (lab)}_{\lambda\mu}$
between the highest-weight states
are given by a relation \cite{Mar76,Mar77}
\begin{equation}
\bra{I_f I_f} Q^{\rm (lab)}_{\lambda \Delta I} \ket{I_i I_i} 
=
(I_f| \tilde{Q}_{\lambda\Delta I} |I_i) ,
\label{high_spin_formula}
\end{equation}
where the state $|I)$ is a symmetry-broken state;
for instance, a mean-field solution of the
cranking Hamiltonian~(\ref{cranking_hamiltonian_1D}) with
the constraint $(I| J_x |I)=I$.
$\tilde{Q}_{\lambda \Delta I}$ can be expanded in terms of
those defined with respect to intrinsic $z$ axis,
and the coefficients are given by the $d$ functions.
\begin{equation}
\tilde{Q}_{\lambda\mu}=
\sum_\nu {\cal D}^\lambda_{\mu \nu}
\left( -\frac{\pi}{2}, -\frac{\pi}{2},0\right)
 Q_{\lambda \nu} 
= i^{-\mu} \sum_\nu d^\lambda_{\mu \nu}\left(-\frac{\pi}{2}\right)
 Q_{\lambda \nu} .
\label{mu_to_nu}
\end{equation}
Now, let us do something not entirely correct.
Although the equality in equation~(\ref{high_spin_formula})
 holds only at high spin,
we take the opposite low-spin limit ($I,\omega_{\rm rot}\rightarrow 0$),
in which the state $|I)$ becomes a ``non-cranked''
$K$-good intrinsic state $|K)$.
Substituting equation~(\ref{mu_to_nu}) into~(\ref{high_spin_formula}),
we have
\begin{equation}
\bra{I_f I_f} Q^{\rm (lab)}_{\lambda \Delta I} \ket{I_i I_i}_{\rm LO}
\leftrightarrow (K_f| \tilde{Q}_{\lambda\Delta I} |K_i) 
= i^{-\Delta I} d^\lambda_{\Delta I \Delta K}
 (K_f| Q_{\lambda \Delta K} |K_i) .
\label{low_spin_formula}
\end{equation}
Here, we use the symbol $\leftrightarrow$ instead of $=$ because
it is obtained by applying
the high-spin formula~(\ref{high_spin_formula}) to the low-spin limit.
For simplicity, we omit the argument $(-\pi/2)$ of the $d$ function.
The suffix ``LO'' indicates the relation in the zeroth order
$O(\omega_{\rm rot}^0)$, with respect to $\omega_{\rm rot}$.

Equation~(\ref{low_spin_formula}) is, of course,
not directly applicable to low spin.
However, it has a proper correspondence to the
leading order (LO) intensity relation in the unified model,
\begin{equation}
\bra{I_f I_f} Q^{\rm (lab)}_{\lambda \Delta I} \ket{I_i I_i}_{\rm LO}
= \bra{K_f I_f I_f} \mathcal{D}^\lambda_{\Delta I \Delta K}
\ket{K_i I_i I_i}
 (K_f| Q_{\lambda \Delta K} |K_i) ,
\label{LO_unified_model}
\end{equation}
which is obtained using the $K$-good wave function~(\ref{unified_model_wf})
and the LO transformation of the multipole operator,
$Q_{\lambda \mu}^{\rm (lab)} =
\sum_\nu \mathcal{D}^\lambda_{\mu\nu} Q_{\lambda\nu}$.
Comparing equations~(\ref{low_spin_formula}) and (\ref{LO_unified_model}),
we may think of a quantization prescription,
\begin{equation}
d^\lambda_{\Delta I\Delta K}
\quad
\rightarrow
\quad
\bra{K_f I_f I_f} {\cal D}^\lambda_{\Delta I\Delta K} \ket{K_i I_i I_i}
.
\label{LO_quantization}
\end{equation}
Then, the ``non-cranked'' limit of the cranking formula
reproduces the LO intensity relation in the unified model.
This quantization procedure is supported by the fact that
the quantities in both sides of equation~(\ref{LO_quantization}) become
identical to the Clebsch-Gordan (CG) coefficients,
$\inproduct{I_i K_i \lambda \Delta K}{I_f K_f}$,
at the high-spin limit ($I\rightarrow\infty$).
Decreasing $I$, the left hand side of equation~(\ref{LO_quantization})
is losing its validity because of its classical nature,
while the right hand side stays valid keeping its quantum nature.

The present quantization of the cranking model is applicable to
higher-order Coriolis coupling terms.
These terms are not easily provided in the unified model.
The next leading order (NLO) is given by the first order in $\omega_{\rm rot}$,
which produces non-zero contributions of $Q_{\lambda \nu=\Delta K\pm 1}$.
The NLO terms to equation~(\ref{low_spin_formula}) are given as
\begin{eqnarray}
\bra{I_f I_f}Q^{\rm (lab)}_{\lambda\Delta I}\ket{I_i I_i}_{\rm NLO}
&\leftrightarrow&
i^{-\Delta I} \omega_{\rm rot} \left( d^\lambda_{\Delta I\ \Delta K+1}
 \frac{d (K_f| Q_{\lambda\ \Delta K+1} |K_i)}{d\omega_{\rm rot}} 
\right.
\nonumber \\
&& \quad\quad\quad 
+ \left. d^\lambda_{\Delta I\ \Delta K-1}
 \frac{d (K_f| Q_{\lambda\ \Delta K-1} |K_i)}{d\omega_{\rm rot}}
 \right)
,
\end{eqnarray}
where the derivatives are evaluated at $\omega_{\rm rot}=0$.
A prescription of the NLO quantization is given by
\begin{equation}
\omega_{\rm rot}
d^\lambda_{\Delta I\ \Delta K\pm 1}
\quad\rightarrow\quad
\bra{K_f I_f I_f} \frac{1}{2 \mathcal{J}}
\left\{ I_\pm, {\cal D}^\lambda_{\Delta I\ \Delta K\pm 1} \right\}
 \ket{K_i I_i I_i} ,
\label{NLO_quantization}
\end{equation}
where $\{A,B\}=AB+BA$ and
$I_\pm\equiv \mp (I_x \pm iI_y)/\sqrt{2}$ in the intrinsic frame.
${\cal J}$ is the moment of inertia of the rotational band,
which can be also calculated in the cranking model
at $\omega_{\rm rot}\rightarrow 0$:
$\mathcal{J} = (1/2)(
d(K_i|J_x|K_i)/d\omega_{\rm rot} +d(K_f|J_x|K_f)/d\omega_{\rm rot})
\approx d(K_i|J_x|K_i)/d\omega_{\rm rot}
\approx d(K_f|J_x|K_f)/d\omega_{\rm rot}$.
Again, in the high-spin limit, the left and right hand sides of
equation~(\ref{NLO_quantization}) become identical,
if we assume $\omega_{\rm rot}\approx I_i/\mathcal{J}\approx I_f/\mathcal{J}$.

In summary, the generalized intensity relation up to the NLO
is obtained by calculating the matrix element
$\bra{I_f I_f} Q^{\rm (lab)}_{\lambda\Delta I} \ket{I_i I_i}=
\bra{K_f I_f I_f} Q^{({\rm LO+NLO})}_{\lambda\Delta I} \ket{K_i I_i I_i}$,
using the operator
\begin{equation}
\fl
Q^{\rm (LO+NLO)}_{\lambda\Delta I}
= m_{\lambda\ \Delta K}^{(0)}
\mathcal{D}^\lambda_{\Delta I \Delta K}
 + \frac{m_{\lambda\ \Delta K+1}^{(+1)}}{2}
\left\{ I_+, {\cal D}^\lambda_{\Delta I\ \Delta K+1} \right\}
+\frac{m_{\lambda\ \Delta K-1}^{(-1)}}{2}
\left\{ I_-, {\cal D}^\lambda_{\Delta I\ \Delta K-1} \right\} ,
\label{LO+NLO}
\end{equation}
where the intrinsic matrix elements are given by
\begin{equation}
m_{\lambda\ \Delta K}^{(0)}
= (K_f| Q_{\lambda \Delta K} |K_i) , \qquad
m_{\lambda\ \Delta K\pm 1}^{(\pm 1)}
= \frac{1}{\mathcal{J}}
\frac{d (K_f| Q_{\lambda\ \Delta K\pm 1} |K_i)}{d\omega_{\rm rot}}  .
\label{intrinsic_moments_LO+NLO}
\end{equation}
The right hand sides of these equations can be calculated
with the cranking model~(\ref{cranking_hamiltonian_1D})
in the vicinity of $\omega_{\rm rot}\rightarrow 0$.
Note that the $\mathcal{R}$-conjugate terms should be added
in the right hand side of equation~(\ref{LO+NLO})
when the $\mathcal{R}$ invariance is present \cite{SN96}.

\subsubsection{Applications}
\label{sec:GIR_applications}

The cranking model~(\ref{cranking_hamiltonian_1D})
has been applied to calculation of the intrinsic moments
in equation~(\ref{intrinsic_moments_LO+NLO}).
For low-lying quadrupole vibrational excitations,
we use the QRPA to calculate the intrinsic matrix elements.
For even-even nuclei, the ground state is $|0)=|K=0)$
and the vibrational state is given by the QRPA normal-mode creation
operator $\hat{X}_K^\dagger$ as $|K)=\hat{X}_K^\dagger |0)$.
The QRPA calculation is based on the cranked-Nilsson-BCS model
with a residual multipole interaction of a separable form similar to
equation~(\ref{separable_interaction}).
We reported these results for quadrupole and
octupole vibrations in the even-even rare-earth nuclei
in reference~\cite{SN96} in which the details of the model can be found.

\begin{figure}[tb]
\centerline{
\includegraphics[clip,width=0.85\textwidth]{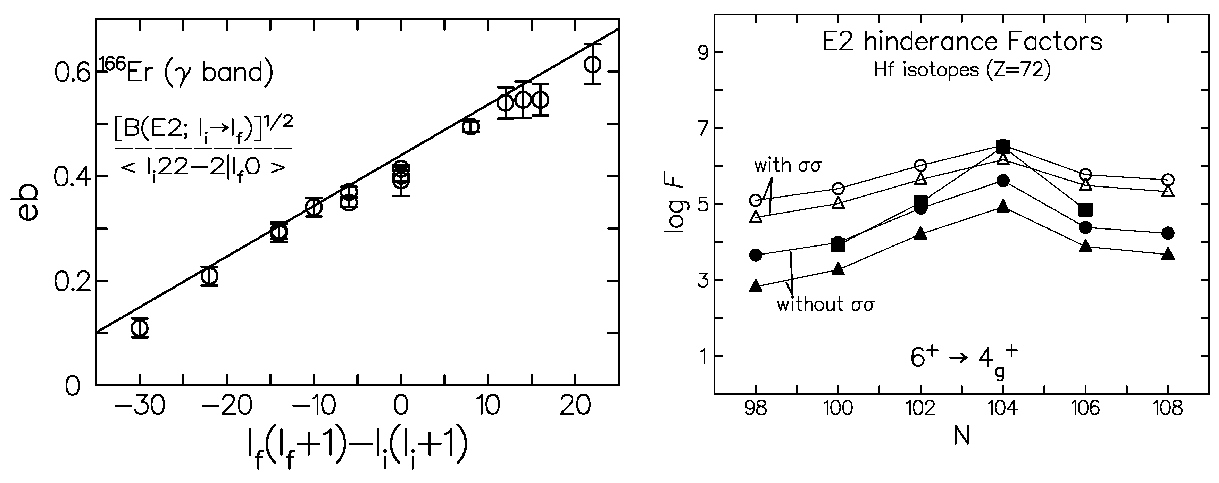}
}
\caption{
(Left) $E2$ transition amplitudes for the $\gamma$ vibrational
band in $^{166}$Er.
The experimental data are taken from figure 4-30 in BM2.
Adapted from reference \cite{NS99-P}.
(Right)
Hindrance factors of $B(E2;6^+_{K=6} \rightarrow 4^+)$
for decay of $K^\pi=6^+$ isomers in Hf isotopes.
Calculated values are shown by circles, while
the squares are the experimental data.
See text for details.
Adapted from reference \cite{NS99-P}.
}
\label{fig:Er_Hf}
\end{figure}

In the left panel of figure~\ref{fig:Er_Hf},
we present an example of the Mikhailov plot
for the $\gamma$ vibrations in $^{166}$Er.
The LO+NLO electric quadrupole operator in a form
of equation~(\ref{LO+NLO}) leads to the intensity relation
between the $K_i=2$ ($\gamma$) band and the $K_f=0$ (ground) band,
\begin{equation}
\fl\qquad
\frac{\bra{I_f K_f^\pi=0^+_g}|M(E2)|\ket{I_i K_i^\pi=2^+_\gamma}}
{\sqrt{2I_i+1}\inproduct{I_i 2 2 -2}{I_f 0}} = 
Q_t \left[ 1+ q \left\{ I_f(I_f+1)-I_i(I_i+1) \right\} \right] ,\quad
\label{GIR}
\end{equation}
where $Q_t$ and $q$ are obtained from the intrinsic moments
(\ref{intrinsic_moments_LO+NLO}),
though some modification is necessary 
because of the $\mathcal{R}$ invariance.
See reference~\cite{SN96} for their exact formulae.

\begin{figure}[htb]
\begin{center}
\includegraphics[clip,width=0.48\textwidth]{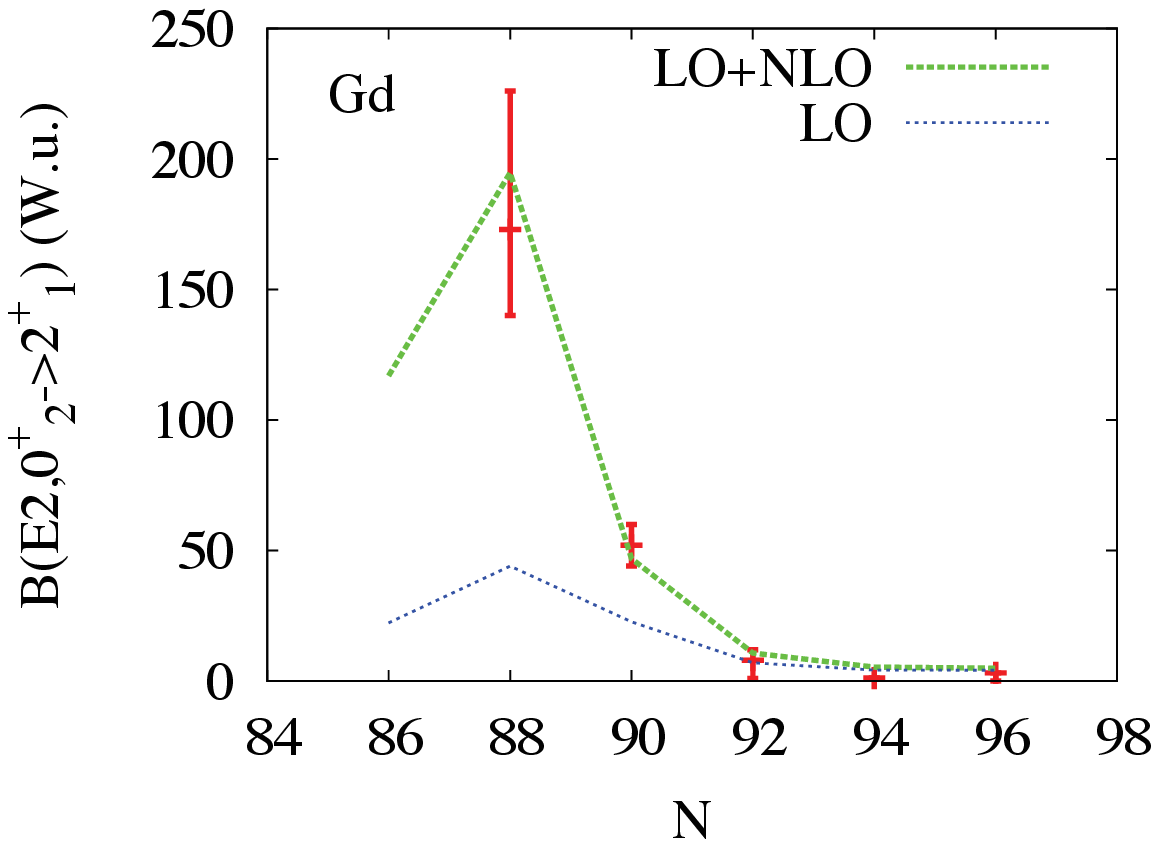}
\includegraphics[clip,width=0.46\textwidth]{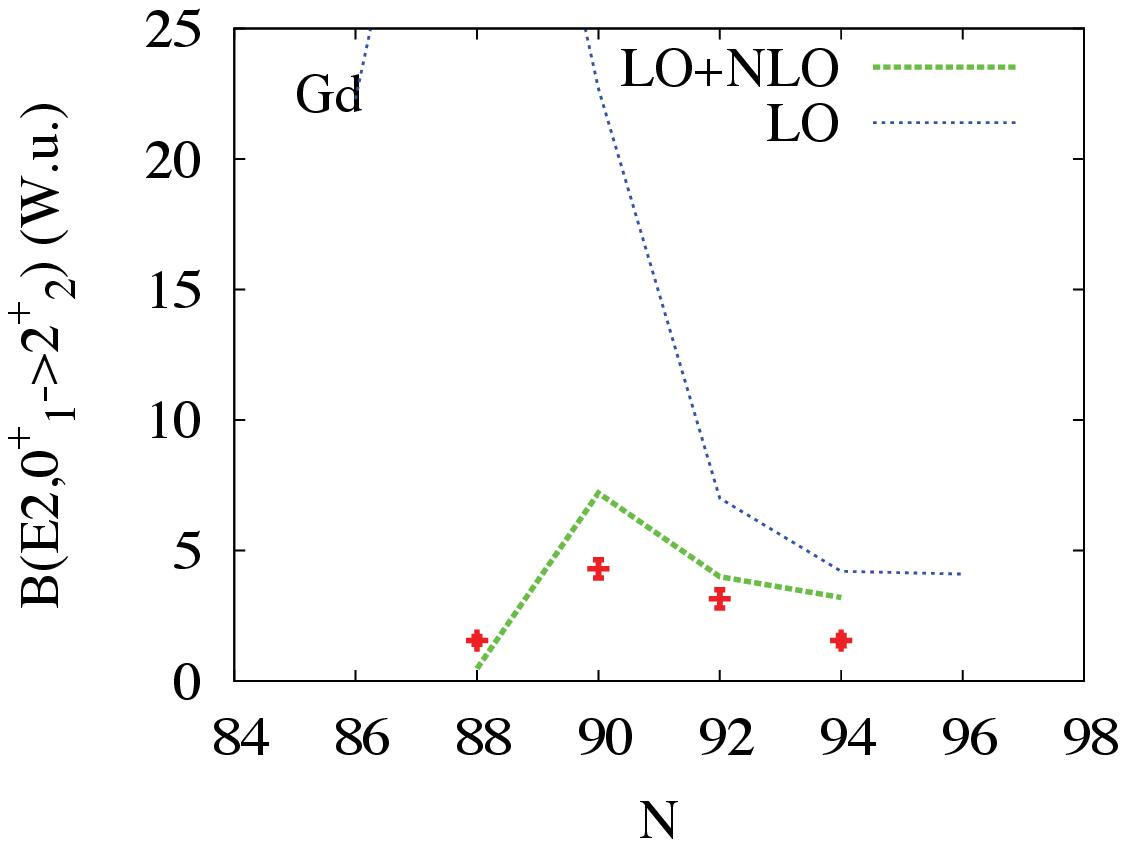}
\end{center}
\caption{
(Left) Calculated and experimental $B(E2;0^+_\beta\rightarrow 2^+_g)$
values in Gd isotopes.
Adapted from reference~\cite{MU16}.
(Right) $B(E2;0^+_g\rightarrow 2^+_\beta)$ values in Gd isotopes.
Experimental data are taken from
references~\cite{Gar01,Mar13,Rei09,Rei12,Hel04,Rei05}.
Note that the scale of the ordinate is 1/10 of that in
the left panel.
}
\label{fig:beta}
\end{figure}

A very similar figure is shown in Fig. 4-30 of BM2,
however, we note here that the parameters $(Q_t,q)$
in the left panel of figure~\ref{fig:Er_Hf}
is based on the microscopic calculation, while those in BM2 are
determined by fitting the experimental data.
The LO relation produces the same $Q_t$ but $q=0$.
The Coriolis coupling effect in the NLO is represented
by the parameter $q$ ($m^{(\pm 1)}_{2\ \mp 1}$).
For the $\gamma$ vibrations, the cranking calculation always
produces $q>0$ in the rare-earth nuclei \cite{SN96}, which suggests
the transitions $I_\gamma \rightarrow I_g$ are enhanced (hindered)
for $I_g>I_\gamma$ ($I_g < I_\gamma$).
This is consistent with experimental data (see reference \cite{SN96} and
references therein).
We also obtain a reasonable agreement for the $M1$ transitions.
However, the calculated sign of the $M1/E2$ mixing amplitudes changes
from nucleus to nucleus, while the observed values are always negative 
\cite{SN96}.

The collectivity of the $\beta$ vibrations measured by the strengths of the
interband transitions to/from the ground band is
weaker than that of the $\gamma$ vibrations in most cases.
However, in rare-earth nuclei in (near) the transitional region,
the $\beta$ vibrations produce
very low excitation energies and
large $B(E2;0^+_\beta\rightarrow 2^+_g)$ values.
In Gd isotopes, for example, their excitation energies are
681 keV and 615 keV for $^{154}$Gd ($N=90$) and $^{152}$Gd ($N=88$),
respectively.
The $B(E2;0^+_\beta\rightarrow 2^+_g)$ values
amount to $52\pm 8$ and $178^{+53}_{-33}$ W.u. \cite{Rei09,Mar13}.
We expect similar values of $B(E2;0^+_g\rightarrow 2^+_\beta)$,
which are predicted to be identical to
$B(E2;0^+_\beta\rightarrow 2^+_g)$ in the LO relation.
Surprisingly, the observed $B(E2;0^+_g\rightarrow 2^+_\beta)$ values are
much smaller than $B(E2;0^+_\beta\rightarrow 2^+_g)$ \cite{Gar01,Rei09,Mar13}.

Figure~\ref{fig:beta} shows the calculated $B(E2)$ values using the
the LO+NLO intensity relation identical to equation (\ref{GIR})
with some trivial changes in the left hand side
($K_i^\pi=2_\gamma^+ \rightarrow K_i^\pi=0_\beta^+$,
$\inproduct{I_i 2 2 -2}{I_f 0}\rightarrow
\inproduct{I_i 0 2 0}{I_f 0}$).
The LO relation cannot account at all for
both large
$B(E2;0^+_\beta\rightarrow 2^+_g)$ and
small $B(E2;0^+_g\rightarrow 2^+_\beta)$ values.
Owing to relatively small moments of inertia
($\mathcal{J}$) for these transitional nuclei,
the inclusion of the NLO terms with large values of $q$
nicely reproduces both of them.
In BM2 (pp.168--175), the band mixing
between the ground and the $\beta$ bands in $^{174}$Hf are presented
to explain the observed intensity relations.
An effect of hindrance of the shape fluctuation
induced by the rotation,
suggested in references \cite{HNMM09,HSNMM10,HSYNMM11},
may also play an important role.
The Coriolis coupling effects may be a key ingredient
to understand the peculiar $B(E2)$ properties of the $\beta$-vibrational bands.

\begin{figure}[tb]
\centerline{\includegraphics[clip,width=0.7\textwidth]{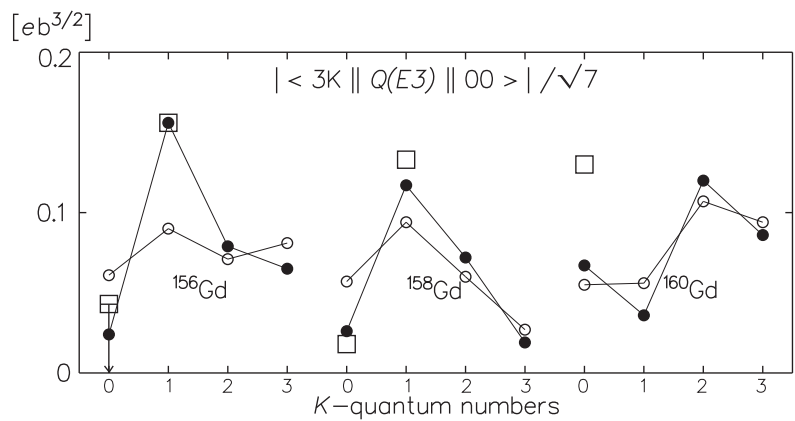}}
\caption{
$E3$ transition amplitudes for the octupole vibrational states
in Gd isotopes, $I_i^\pi=3^- \rightarrow I_f^\pi=0^+$.
Open and filled circles correspond to calculated values
in the LO and LO+NLO, respectively,
compared with experimental data (open squares)
\cite{MM81}.
The lowest mode of excitation among the octupole multiplet is
$K=1$ for $^{156,158}$Gd and $K=2$ for $^{160}$Gd.
From reference \cite{NS97-P}.
}
\label{fig:Gd}
\end{figure}

Generally speaking, the Coriolis-coupling effect for the quadrupole
vibrations is relatively weak, because the low-lying 
$K^\pi=1^+$ collective state is missing
(section~\ref{sec:beta_gamma_vib}).
In contrast, all the members of the multiplet are present for
the negative-parity octupole vibrations.
Thus, we expect stronger Coriolis effects.
The cranking calculation actually predict the NLO parameters
of the octupole vibrations
($|q|\sim 0.1$) larger than those of $\gamma$ vibrations ($|q|\sim 0.01$)
\cite{SN96}.
The $K$ quantum number of the lowest mode of excitation
among the octupole multiplet ($K^\pi=0^-\sim 3^-$)
changes from nucleus to nucleus.
Nevertheless, the lowest mode always has $q<0$ for transitions from
the octupole band ($K_i$) to the ground band ($K_f=0$),
Thus, $B(E3; 3^-\rightarrow 0^+)$ is enhanced for the lowest mode.
This Coriolis effect is clearly seen in Gd isotopes in figure~\ref{fig:Gd}.

Another application is presented here for $K$-forbidden transitions
in decays of the high-$K$ isomers.
For $K$-forbidden transitions with $|\Delta K| > \lambda$,
we should extend the LO+NLO relation of equation~(\ref{LO+NLO})
because the LO term vanishes.
The order of $K$ forbiddenness is defined by $n=|\Delta K|-\lambda$.
The N$^n$LO+N$^{n+1}$LO intensity relation for $\Delta K > 0$ is given by
\cite{SN96}
\begin{equation}
\fl\qquad
 Q^{{\rm (N}^n+{\rm N}^{n+1}{\rm LO)}}_{\lambda\Delta I}
= m_{\lambda\ \lambda}^{(-n)}
\mathcal{D}^\lambda_{\Delta I \lambda} (I_-)^n
+\frac{m_{\lambda\ \lambda-1}^{(-n-1)}}{2}
\left\{ (I_-)^{n+1}, {\cal D}^\lambda_{\Delta I\ \lambda-1} \right\}
+ \mathcal{R}{\rm -conj.} ,\quad
\end{equation}
where the intrinsic moments are
\begin{equation}
\eqalign{
m_{\lambda\ \pm\lambda}^{(\mp n)}
= \frac{1}{n! \mathcal{J}^n}
\frac{d^n (K_f| Q_{\lambda\ \pm\lambda} |K_i)}{d\omega_{\rm rot}^n}  ,\cr
m_{\lambda\ \pm(\lambda-1)}^{(\mp(n+1))}
= \frac{1}{(n+1)! \mathcal{J}^{n+1}}
\frac{d^{n+1} (K_f| Q_{\lambda\ \pm(\lambda-1)} |K_i)}{d\omega_{\rm rot}^{n+1}}
.
}
\end{equation}
These formulae are applied to the two-quasiparticle (2qp)
$K^\pi=6^+$ isomers in Hf isotopes.
The configuration
of the initial state $|K_i=6)$
is assumed to be the proton 2qp
$[402\ 5/2]\otimes[404\ 7/2]$.
The hindrance factors are shown in the right panel of figure~\ref{fig:Er_Hf}.
This is defined by
\begin{equation}
F\equiv \frac{B(E2;6^+\rightarrow 4^+)_{W}}{B(E2;6^+_{K=6} \rightarrow 4^+_g)}
,
\end{equation}
where $B(E2)_W$ is the Weisskopf estimate of the reduced transition probability.
A large hindrance factor means a long life time of the high-$K$
isomer state.

The calculated values are shown 
in the right panel of figure~\ref{fig:Er_Hf}
by filled symbols (circles and triangles),
which are compared with experiment (filled squares).
The calculation qualitatively reproduces the experimental trends.
However, these values turn out to be quite sensitive to the details of
the quasiparticle spectra.
For instance, the triangles are obtained with slightly larger values of
the proton chemical potentials (by 70 keV) than those of circles.
The calculated hindrance factors differ by about one order of magnitude.
The effect of the residual interaction is very significant too.
The open symbols in figure~\ref{fig:Er_Hf} (right) show results including
the spin-spin interaction, $V_0\vec{\sigma}\cdot\vec{\sigma}$,
with $V_0=100$ keV, which roughly accounts for the Gallagher-Moszkowski
splitting.
This could change $F$ by two order of magnitude.
Nonetheless, the neutron number dependence is rather universal,
that indicates the largest hindrance at $N=104$.

\subsection{Cranking model at high spin: QRPA in the rotating frame}
\label{sec:cranking_at_high_spin}

Rotating the nucleus very fast,
the perturbative treatment in section~\ref{sec:GIR}
becomes no longer valid.
Instead, the semiclassical treatment in the cranking model
is better justifiable, and
the direct application of the cranking model~(\ref{cranking_hamiltonian_1D})
has been extensively performed for a variety of high-spin phenomena.
For instance, the famous back-bending phenomena have been studied
and understood as the crossing between the ground band and
an aligned 2qp band.
This can be also interpreted as breaking a Cooper pair
condensed in the ground band by the Coriolis anti-pairing effect.

In order to investigate properties of elementary modes of excitation
at high spin, it is useful to use the ``routhian'' $E'(\omega_{\rm rot})$
as a function of the rotational frequency $\omega_{\rm rot}$.
The routhian here is defined as the eigenenergies of the
cranking Hamiltonian~(\ref{cranking_hamiltonian_1D}),
which can be interpreted as the energy in the rotating frame
with the rotational frequency $\omega_{\rm rot}$.
To make a comparison, we often convert
the experimental excitation energy as a function of $I$, $E(I)$,
into the ``routhian'' $E'(\omega_{\rm rot})$.
This is done as follows.
First, from experimental rotational spectra $E_b(I)$,
we calculate the frequency, $\omega_{\rm rot}(I)=dE_b(I)/dI$.
Here, $b$ is the index of the rotational band.
According to the cranking Hamiltonian~(\ref{cranking_hamiltonian_1D}),
the routhian is defined as
$E'_b(\omega_{\rm rot})=E_b(I)-\omega_{\rm rot}I_x(I)$
with $I_x(I)=\sqrt{(I+1/2)^2-K^2}$,
at discrete values of $\omega_{\rm rot}(I)$.
The reference routhian $E'_{\rm ref}(\omega_{\rm rot})$ is defined,
for instance, by fitting that of the ground-state band
(``{\it b}''=``g.s.'').
Then, the excitation routhian relative to the reference band
as a function of $\omega_{\rm rot}$ is obtained
as $E'_{\rm ex}(\omega_{\rm rot})=E'_b(\omega_{\rm rot}) - 
E'_{\rm ref}(\omega_{\rm rot})$ for each band ``$b$''.
In Ref.~\cite{BF79}, the experimental routhians in odd nuclei,
which were obtained by adopting the reference band fitting the
ground-state band in neighboring even-even nuclei,
show nice agreement with the calculated quasiparticle routhians.

\begin{figure}[tb]
\centerline{
\includegraphics[clip,width=0.45\textwidth]{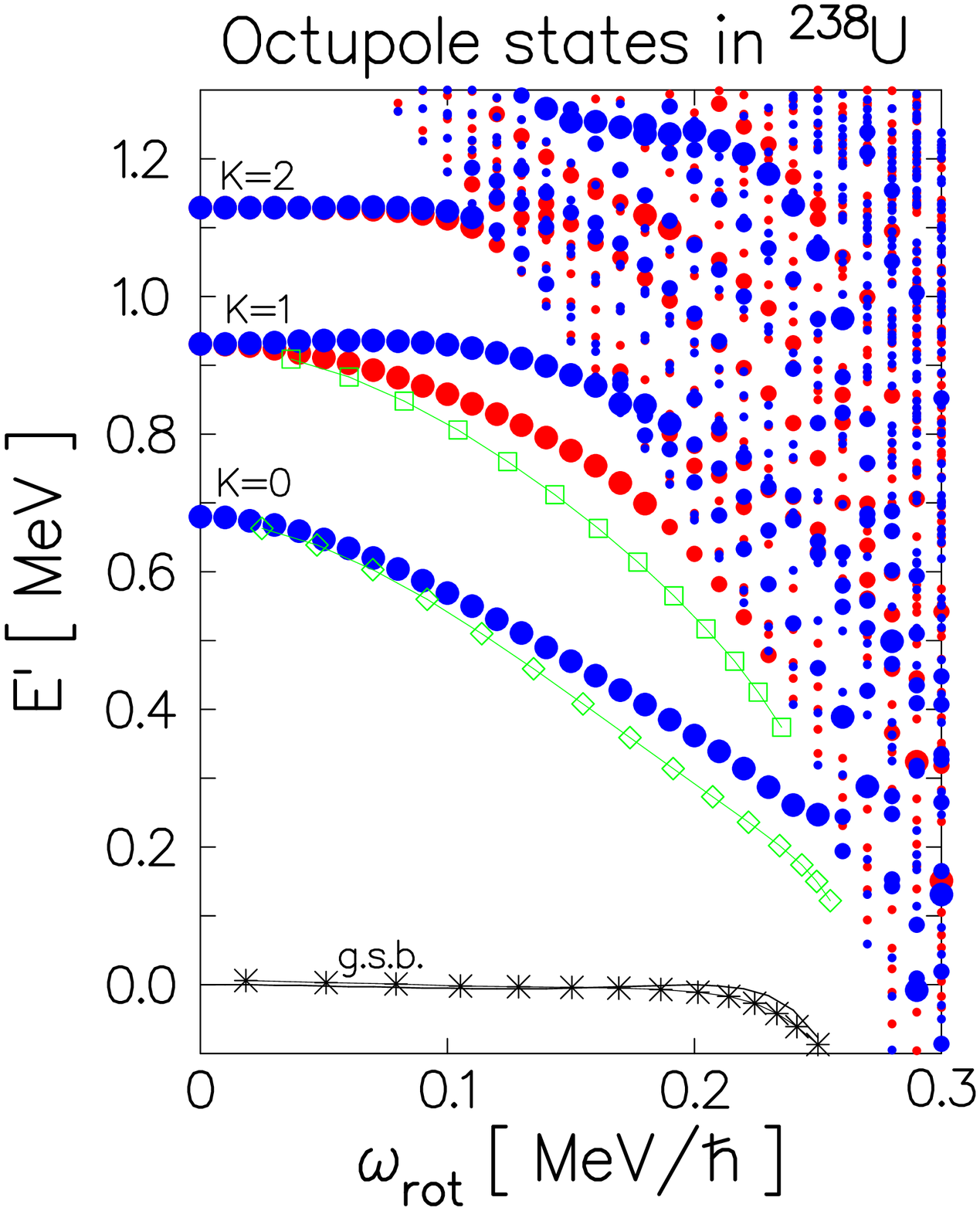}
\includegraphics[clip,width=0.4\textwidth]{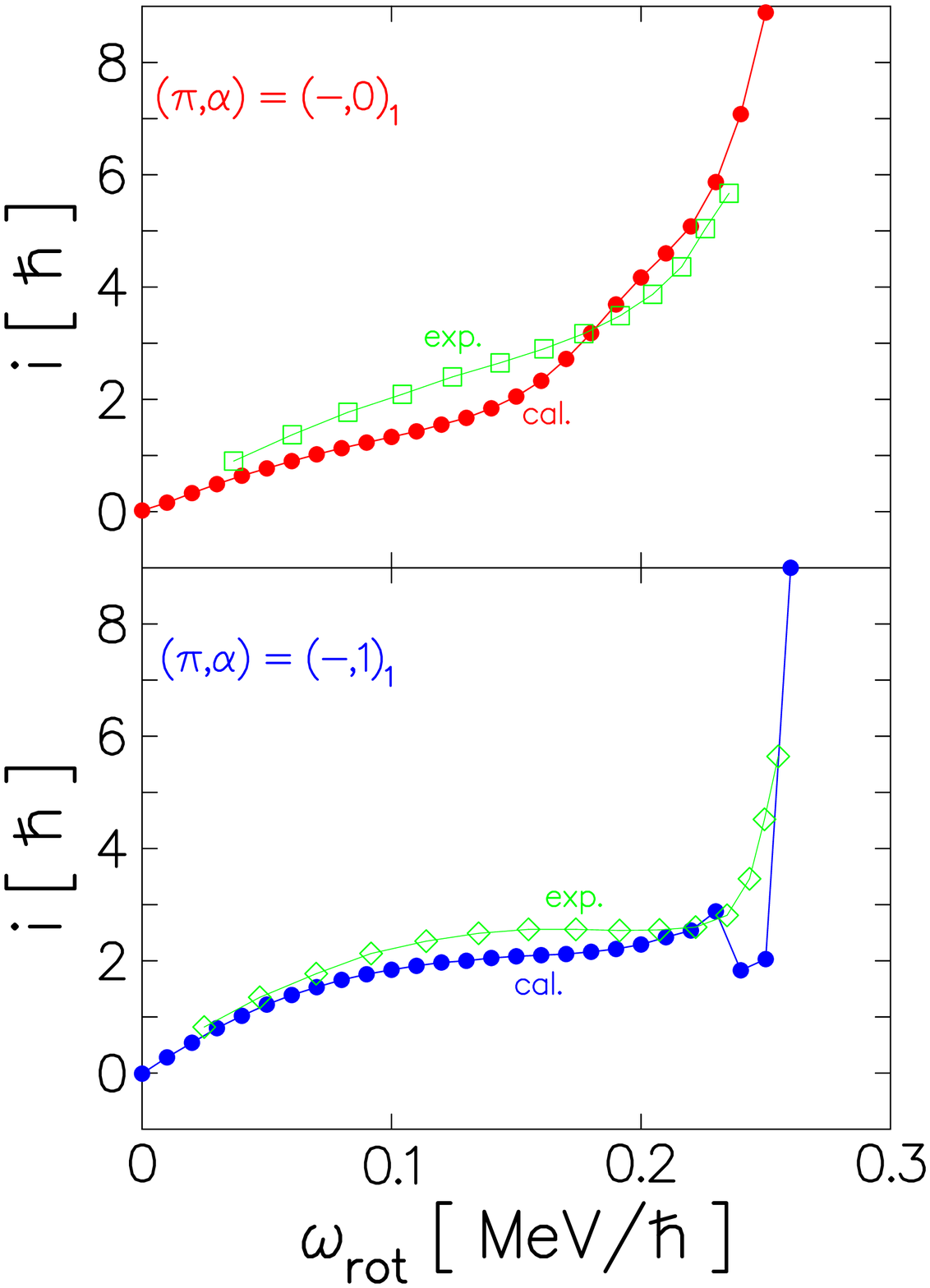}
}
\vspace{-15pt}
\caption{(Left)
Excitation routhian plot as functions of rotational frequency
$\omega_{\rm rot}$ for negative-parity states in $^{238}$U.
Large, medium, and small circles indicate the QRPA solutions with
$E3$ transition amplitudes larger than 300 $e$fm$^3$,
larger than 150 $e$fm$^3$, and less than 150 $e$fm$^3$, respectively.
The red (blue) ones correspond to the signature even (odd) states
with even (odd) $I$.
Experimental routhians are plotted by open squares.
(Right)
Aligned angular momentum as a function of $\omega_{\rm rot}$
for the lowest ($K^\pi=0^-$) and the second lowest ($K^\pi=1^-$)
octupole bands
in the lower and upper panels, respectively.
Open squares indicate experimental data \cite{War96}.
Adapted from reference \cite{Nak96-P}.
}
\label{fig:U}
\end{figure}

The routhian plot for the octupole vibrational bands
in $^{238}$U are presented in figure~\ref{fig:U}.
See references~\cite{NMMS96,Nak96-P} for details of the calculation.
In the right panels, the alignment, defined by
$i\equiv -dE'/d\omega_{\rm rot}$, is shown.
The alignment indicates the aligned component of the angular
momentum carried by the vibrational excitation.
For the lowest octupole band with $K^\pi=0^-$,
the alignment quickly increases up to $i\sim 3$,
which suggests that the angular momentum of the octupole phonon
is almost fully aligned along the rotational axis.
Then, at high spin around $\omega=0.25$ MeV,
it suddenly jumps up, which suggests the breakdown of the
collective vibration by a strong Coriolis force.
At $\omega_{\rm rot}\gtrsim 0.25$ MeV, 
it becomes an aligned 2qp state.
This is seen in the left panel too.
The octupole collectivity (size of the circles) suddenly decreases
around $\omega_{\rm rot}=0.25$ MeV.
In contrast to the lowest band, the second lowest $K^\pi=1^-$ band
with even $I$ shows a gradual increase in the alignment, which may
suggest the gradual change of the octupole phonon into the aligned
2qp structure.
The present calculation nicely agrees with the experimental data.

The argument here suggests that the vibrational excitations based
on the yrast (ground-state) band tend to lose their collective character
at high spin, due to the intrusion of the aligned 2qp
states at low energy.
In this respect, the nucleus with larger deformation may be
better suited for the observation of the rapidly rotating
vibrational bands.
This is because the large deformation tends to hinder the alignment
of the quasiparticles.
Next, let us discuss such a case, the high-spin superdeformed (SD) bands.

\section{Elementary excitations in superdeformed rotational bands}
\label{sec:SD}

The SD state is characterized by a large prolate
deformation of approximate two-to-one axis ratio.
The shell structure at the SD shape is very different
from that near the spherical shape.
Since the low-lying modes of excitation strongly depend on the
underlying shell structure,
we may expect some new features in their properties.

\subsection{Octupole vibrations with $K^\pi=0^-$ and $2^-$}

One of the most striking features in the SD shell
structure is that the single-particle levels with
opposite parity ($\pi=\pm$) coexist in a single shell.
Adopting a simple harmonic oscillator potential,
one can easily understand this fact:
Namely, for $\omega_x=\omega_y=2\omega_z$,
an orbital with the oscillator quanta $(n_x,n_y,n_z)$
is degenerate in energy with those of
$(n_x\mp 1,n_y,n_z\pm 2)$ and $(n_x,n_y\mp 1,n_z\pm 2)$.
Since the parity is determined by the total quanta $N=n_x+n_y+n_z$,
they have different parity.
Another feature is that the observed SD bands are located around
the closed shell configurations corresponding
to the SD magic numbers (See figure 6-50 in BM2).
In contrast, the normally deformed nucleus is a consequence of
the SBS and has an open-shell configuration away from the spherical
magic numbers.
From these simple analysis, we may expect that
the collective negative-parity modes of excitation appear at low energy.

\begin{figure}[tb]
\centerline{
\includegraphics[width=0.8\textwidth]{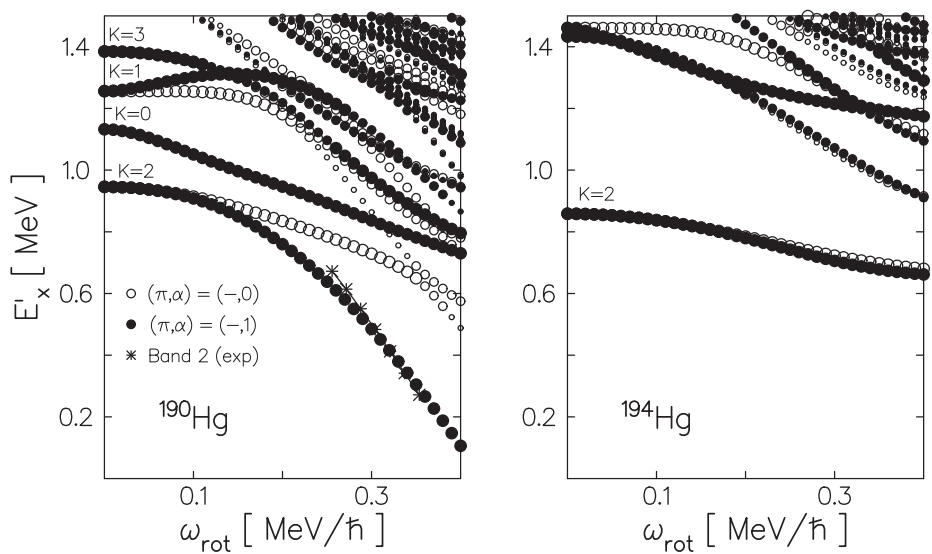}
}
\caption{
Excitation routhian plot for negative-parity excitations in SD $^{190}$Hg
(left) and $^{194}$Hg (right).
Open and filled circles correspond to the states
with even and odd signatures, respectively.
The size of circles represent the $E3$ transition amplitudes
($>200$ $e$fm$^3$, $>100$ $e$fm$^3$, and $<100$ $e$fm$^3$)
Experimental excitation routhians in $^{190}$Hg are shown by stars ($*$).
From reference \cite{Nak96-P}.
}
\label{fig:Hg}
\end{figure}

The QRPA based on the cranked Nilsson-BCS model is applied to
SD bands in the $A=190$ region.
The calculation predicts that the $K^\pi=2^-$ octupole states
are particularly low in energy, around $E'_x\lesssim 1$ MeV.
Especially, in $^{194}$Hg and $^{196}$Pb with $N=114$,
very collective $K^\pi=2^-$ octupole vibrations appear well below
1 MeV and their excitation routhians are roughly constant
with very little signature splitting \cite{NMMS96}.
See the right panel of figure~\ref{fig:Hg}.
Later, the interband $E1$ transitions between the octupole and ground SD
bands have been measured for $^{194}$Hg \cite{Hac97}
and for $^{196}$Pb \cite{Ros01},
which confirms nice agreement with calculated routhians
and the strong octupole collectivity.
For $N=110$, the calculation predicts an aligned octupole phonon,
shown in the left panel of figure~\ref{fig:Hg},
similar to the lowest octupole band in $^{238}$U
in figure~\ref{fig:U}.
This also nicely reproduces the experimental routhians in an
excited SD band in $^{190}$Hg \cite{Wil96}.
Later, the linear polarization measurement confirms the
aligned octupole vibrations \cite{Kor01}.

In the $A=150$ region, we theoretically predicted 
a possible candidate of $K^\pi=0^-$ octupole band in SD
$^{152}$Dy \cite{NMMN95}.
It shows a rather constant excitation routhians in a wide range
of $\omega_{\rm rot}=0\sim 0.7$ MeV.
Later in 2002, its octupole character has been confirmed by the
measurement of the interband transitions and the spin identification
\cite{Lau02-1,Lau02-2}.
The $\omega_{\rm rot}$-dependence of the routhian well agrees
with the theoretical prediction.

\subsection{Soft mode with $K^\pi=1^-$}
\label{sec:banana_SD}

\begin{figure}[tb]
\includegraphics[width=0.3\textwidth]{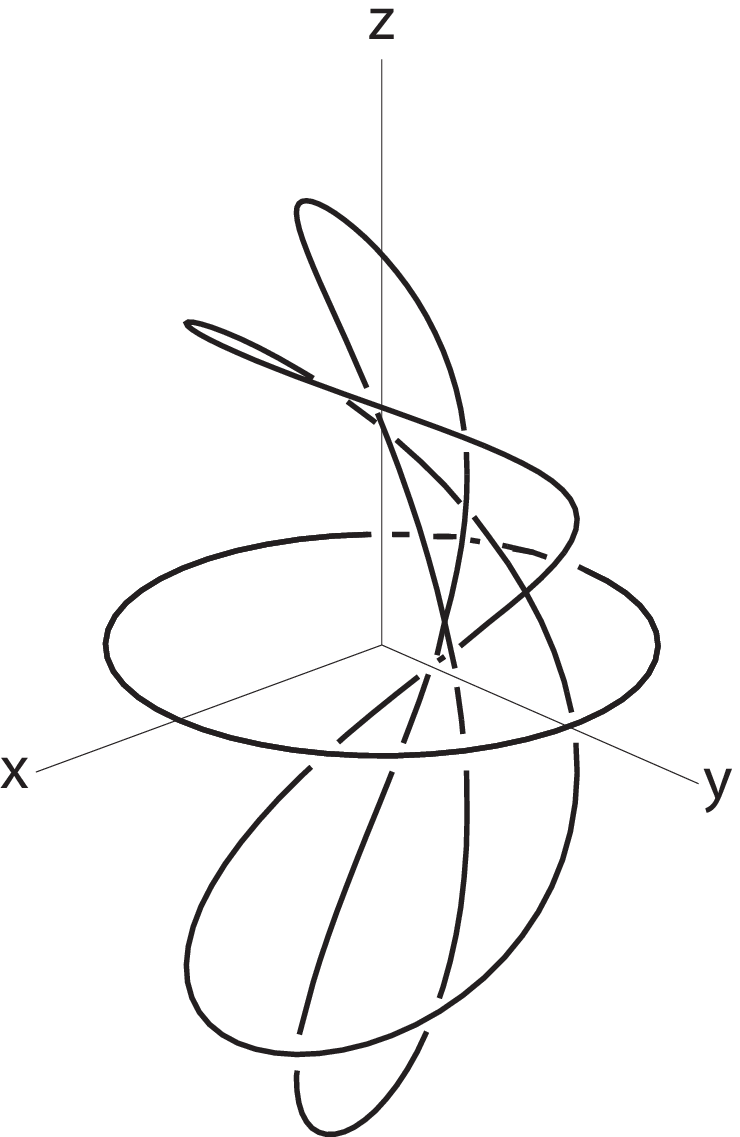}
\includegraphics[clip,width=0.7\textwidth]{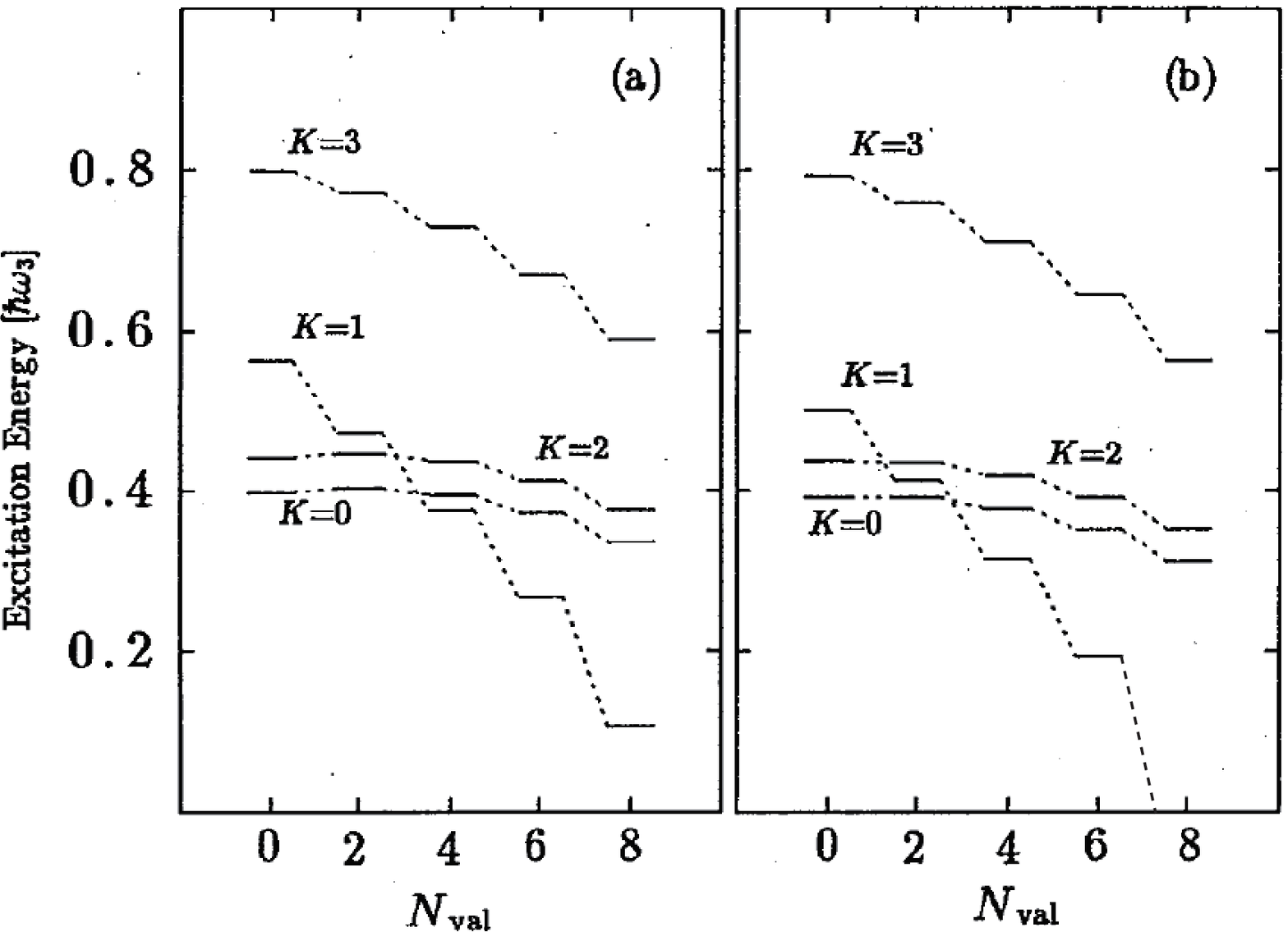}
\caption{
(Left)
Typical classical periodic orbits in a potential with the $a:b:c=2:2:1$
shell structure.
From reference \cite{MNA93-P}.
(Right)
Calculated excitation energies for octupole vibrations in
the SD harmonic oscillator potential, with
stronger pairing (a) and weak pairing interactions (b).
From reference \cite{NMM92}.
}
\label{fig:HO}
\end{figure}

So far, the octupole vibrational excitations in SD rotational bands
have been observed (confirmed) only for the $K^\pi=0^-$ and $2^-$ modes.
Theoretically, these modes are predicted to appear lower than other
modes ($K^\pi=1^-$ and $3^-$), near the SD magic numbers.
However, moving away from the magic closed configurations,
the $K^\pi=1^-$ modes become a soft mode.

In order to investigate the soft mode in the SD shape,
we again follow the discussion in BM2 (pp. 591$-$598),
extending the argument for spherical potentials
in section~\ref{sec:shell_structure} to a deformed one.
This is possible if the motion is separable in the three coordinates,
such as the harmonic oscillator potential.
In a deformed harmonic oscillator potential,
the single-particle energy is specified by there numbers of
the oscillator quanta, $(n_x,n_y,n_z)$.
Thus, the shell structure is characterized by the ratio of
three integers, $a:b:c$,
and the shell frequency given by
\begin{equation}
\fl\qquad
\omega_{\rm sh}\equiv \frac{1}{a}
\left(\frac{\partial\epsilon}{\partial n_x}\right)_0
= \frac{1}{b}
\left(\frac{\partial\epsilon}{\partial n_y}\right)_0
= \frac{1}{c}
\left(\frac{\partial\epsilon}{\partial n_z}\right)_0 .
\end{equation}
Since $a:b:c=1:1:1$ correspond to the spherical harmonic oscillator,
the simplest integer ratio next to $1:1:1$
is $a:b:c=1:1:2$ and $2:2:1$.
The SD shape we are discussing here corresponds to the latter one,
which has the prolate shape.

The frequency ratio of $a:b:c=2:2:1$
($\omega_x=\omega_y=2\omega_z=2\omega_{\rm sh}$)
creates periodic orbits
shown in the left panel of figure~\ref{fig:HO}.
These are orbits of ``bending figure of eight''.
Since the shape of the classical periodic orbits is related to the
soft mode,
the SD state, with many nucleons outside the closed shell,
may be unstable against the banana-shaped bending mode.

In figure~\ref{fig:HO}, we show the result of the QRPA calculation
with the separable octupole interaction, based on
the SD harmonic oscillator potential \cite{NMM92}.
The $K^\pi=0^-$ and $2^-$ modes are the lowest near the SD magic numbers.
These modes are rather insensitive to the number of nucleons outside
the closed shell.
However, the $K^\pi=1^-$ octupole mode dramatically decreases its energy as
increasing
the number of
valence nucleons.
With enough number of valence nucleons,
the bending $K^\pi=1^-$ mode leads to the instability.

According to the qualitative discussion on the SD shell structure,
Bohr and Mottelson have already pointed out the possibility of
this instability toward the bending shape, in the context of
fission path (p.598 in BM2).
As far as we know, this effect on the fission dynamics
has not been fully studied so far.

\section{Nuclear wobbling motion and precession}
\label{sec:wobbling}

Most of the existing experimental data are known to be
consistent with
the interpretation based on the axially symmetric deformation.
Even the octupole deformation (section~\ref{sec:shell_structure})
observed in heavy nuclei
is associated with the axially symmetric one ($Y_{30}$).
In section~\ref{sec:banana_SD}, we have presented
a possible exotic nuclear shape in SD nuclei away from
the closed shell
configuration,
that breaks both the axial and the parity symmetry.
However, it has not been observed in experiments.

Bohr and Mottelson gave extensive discussion on the spectra of
triaxial nuclei in BM2.
In the beginning of section 4-5, they said
``{\it Although, at present {\rm (1975)},
there are no well-established examples of nuclear
spectra corresponding to asymmetric equilibrium shapes, it appears likely
that such spectra will be encountered in the exploration of nuclei
under new conditions (large deformations, angular momentum, isospin, etc.).}''
They were absolutely right.

The identification of the static triaxial deformation
has been a longstanding issue
in the nuclear structure physics.
One of the difficulties is to confirm its ``static'' nature
clearly distinguished from the ``dynamic'' one.
The observation of a wobbling band was
a breakthrough that provided a clear indication
of the non-uniform three-dimensional rotation of a triaxial nucleus.
We have really encountered this new phenomenon at
{\it large deformation and angular momentum}.

The first observation of the wobbling band was in $^{163}$Lu
\cite{Ode01,Jen02}.
At high spin ($I\gtrsim 20$),
several regular rotational bands with large moments of inertia 
come down to the yrast region.
The deformation of these bands have been speculated to be large
($\epsilon\sim 0.4$) and triaxial ($\gamma\sim 20^\circ$),
according to calculations of the total routhian surface (TRS)
with the Nilsson potential~\cite{Gor04}.
They are called the triaxial superdeformed (TSD) bands
and labeled as TSD1, TSD2, etc.
In addition to the stretched intraband transitions ($\Delta I=2$),
the linking transitions ($\Delta I=1$)
between TSD2 and TSD1 (yrast),
between TSD3 and TSD2, and
between TSD3 and TSD1, have been observed.
The $E2$ character of these interband transitions
is experimentally confirmed~\cite{Gor04} and
their large strengths
nicely correspond to the estimate by a simple triaxial rotor model.
The measured $B(E2)$ values for the interband transitions
are order of 100 W.u.
which is
considerably larger than those of
the most collective $\gamma$ vibrations.

\subsection{Rotor model analysis of the wobbling in the high-spin limit}
\label{sec:rotor_model}

\begin{figure}[tb]
\centerline{
\includegraphics[clip,width=0.45\textwidth]{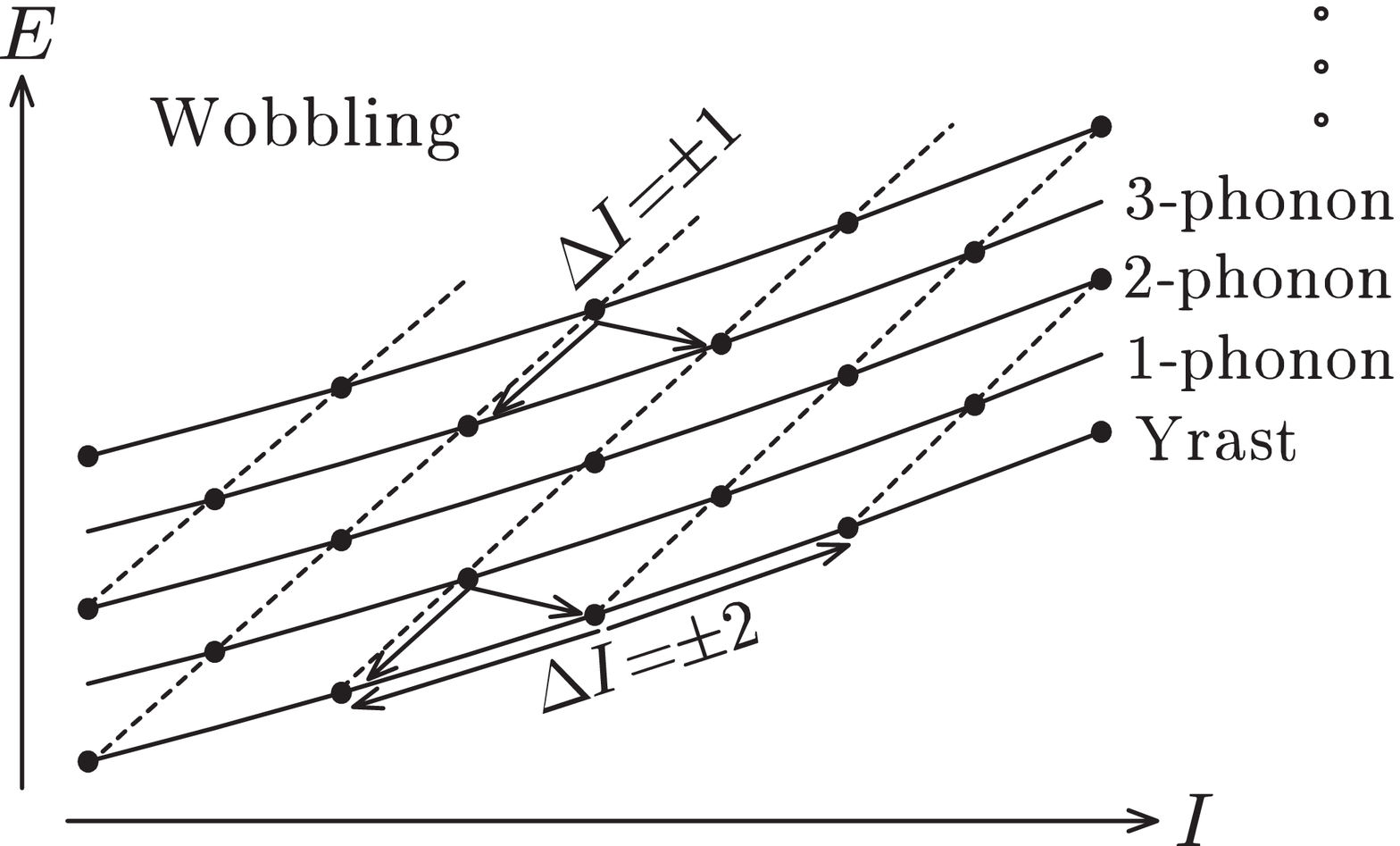}
\hspace*{0.05\textwidth}
\includegraphics[clip,width=0.18\textwidth]{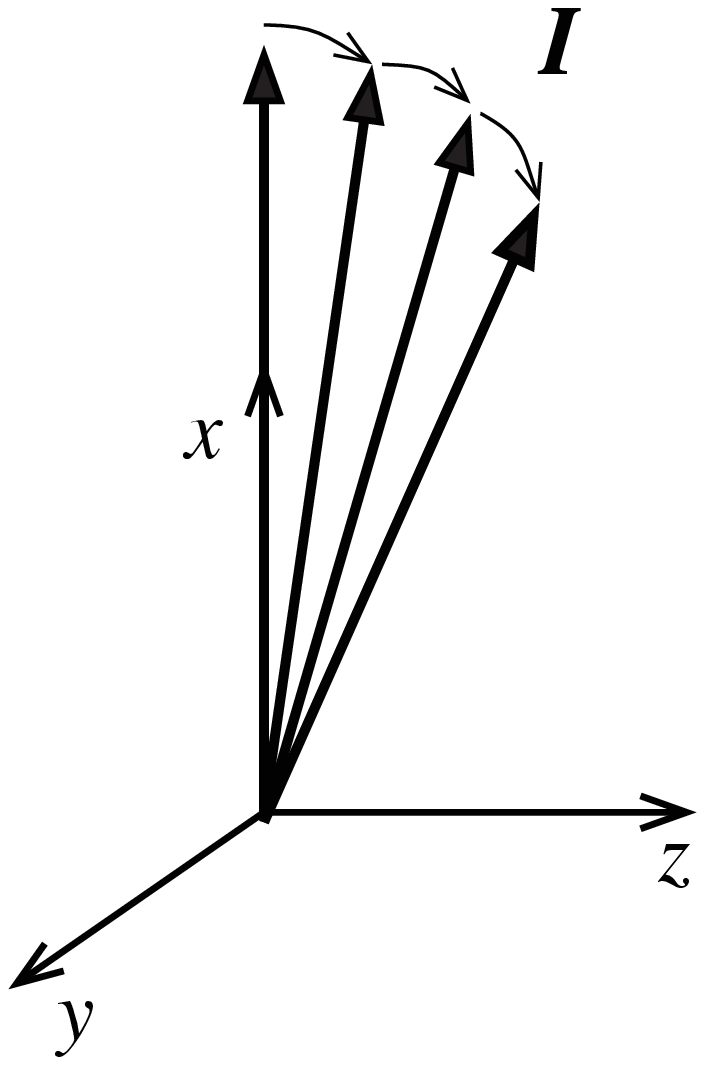}
}
\caption{
A schematic illustration of the wobbling motion.
(Left) Excitation spectra with $\Delta I=2$ and $\Delta I=1$ sequences.
(Right) A wobbling phonon excitation tilts the direction of
the angular momentum from the $x$-axis.
From reference \cite{SS06}.
}
\label{fig:wobbling}
\end{figure}

The prediction based on the rotor model given by Bohr and Mottelson
(section 4-5e in BM2)
is recapitulated here.
The rotor Hamiltonian contains three different moments of inertia,
$\mathcal{J}_x > \mathcal{J}_y > \mathcal{J}_z$,
with respect to the principal axes in the body-fixed frame.
\begin{equation}
\fl\qquad
H_{\rm rot}=\frac{J_x^2}{2\mathcal{J}_x}
+\frac{J_y^2}{2\mathcal{J}_y}
+\frac{J_z^2}{2\mathcal{J}_z} 
=\frac{\vec{J}^2}{2\mathcal{J}_x}
+\left(\frac{1}{2\mathcal{J}_y}-\frac{1}{2\mathcal{J}_x}\right)J_y^2
+\left(\frac{1}{2\mathcal{J}_z}-\frac{1}{2\mathcal{J}_x}\right)J_z^2
.
\label{triaxial_rotor}
\end{equation}
For the lowest energy (yrast) state at a given $I$,
the term proportional to $\vec{J}^2$ in this Hamiltonian
is dominant at high spin ($I\rightarrow\infty$).
This corresponds to a uniform rotation around the $x$ axis:
$E_I\approx I(I+1) /(2\mathcal{J}_x)$.
In this high-spin limit, we assume $J_x\approx I$
which can be treated as a $c$-number.
The remaining terms of the Hamiltonian~(\ref{triaxial_rotor})
can be diagonalized, $[H_{\rm rot},X_{\rm wob}^\dagger]=\omega_{\rm wob}
X_{\rm wob}^\dagger$, by a linear transformation.
The normal-mode (wobbling phonon) creation operator
\begin{equation}
X_{\rm wob}^\dagger \equiv a \frac{iJ_y}{\sqrt{2I}} - b \frac{J_z}{\sqrt{2I}} ,
\end{equation}
with the normalization 
$\left[X_{\rm wob},X_{\rm wob}^\dagger \right] = 1$
leads to the following relations:
\begin{eqnarray}
\frac{a}{b}&=&
\sqrt{
\left(\frac{1}{\mathcal{J}_y}-\frac{1}{\mathcal{J}_x}\right)
\left(\frac{1}{\mathcal{J}_z}-\frac{1}{\mathcal{J}_x}\right)^{-1} } ,
\quad\quad     ab=1 ,
\\
\omega_{\rm wob}
&=& I
\sqrt{
\left(\frac{1}{\mathcal{J}_y}-\frac{1}{\mathcal{J}_x} \right)
\left(\frac{1}{\mathcal{J}_z}-\frac{1}{\mathcal{J}_x} \right) }
.
\label{wobbling_frequency}
\end{eqnarray}
The operator for the wobbling phonon number is given by
$n\equiv X_{\rm wob}^\dagger X_{\rm wob}$.
In this way, the rotor Hamiltonian can be written as a sum of
the rotation around the $x$ axis and the wobbling phonon excitation:
$E_{In}\approx I(I+1)/(2\mathcal{J}_x) + \omega_{\rm wob} (n+1/2)$.
A schematic figure for these spectra is shown in figure~\ref{fig:wobbling}.
In order to realize this kind of multiple band structure
from one intrinsic configuration, the nucleus should be able
to rotate about all three principal axes.
Therefore, the nuclear shape must be triaxial.
In addition, among the three moments of inertia,
the one along the major rotational axis $\mathcal{J}_x$ must be the largest.

According to the LO high-spin formula~(\ref{high_spin_formula}),
the intraband $B(E2;\Delta I=2)$ strengths are proportional to
the quadrupole deformation $(\tilde{Q}_{22})^2$.
The $\Delta I=1$ transitions, which are associated with the wobbling
transitions, appear in the NLO with respect to $1/I$;
$B(E2;\Delta I=1)/B(E2;\Delta I=2) \sim 1/I$.
These $E2$ transition strengths
were explicitly given in chapter 4-5e in BM2.
The TRS calculation predicts the
``positive'' $\gamma$ shape
($\gamma\sim 20^\circ$)
for the TSD bands in $^{163}$Lu.
Here we use the so-called Lund convention for
the triaxiality parameter $\gamma$
in relation to the main rotation axis~\cite{And76},
where for the positive $\gamma$ shape, $0 < \gamma < 60^\circ$,
the rotation ($x$) axis is the shortest principal axis,
while for the negative $\gamma$ shape, $-60^\circ < \gamma < 0$,
it is the intermediate principal axis.
Then the out-of-band $E2$ transitions for the positive $\gamma$ shape
satisfy
a relation, $B(E2;I\rightarrow I-1)> B(E2;I\rightarrow I+1)$.
This is consistent with the experiments, in which only
the $I\rightarrow I-1$ transitions have been observed.

The simple rotor model picture, however, disagrees with
the observed data with respect to the following
points:
\begin{itemize}
\item
At $\gamma=20^\circ$ which is supported by the TRS calculation,
the $\gamma$-dependence of the irrotational
moments of inertia, which are commonly
assumed in the rotor model,
produce $\mathcal{J}_y >\mathcal{J}_x >\mathcal{J}_z$.
This contradicts the basic assumption of
$\mathcal{J}_x >\mathcal{J}_y >\mathcal{J}_z$ and
the formula~(\ref{wobbling_frequency}).
\item
According to equation~(\ref{wobbling_frequency}),
the wobbling frequency $\omega_{\rm wob}$ increases as a function of $I$.
Conversely, the observed frequency decreases.
\end{itemize}
Solutions to these problems will be provided by microscopic treatments
in section~\ref{sec:QRPA_model}.

\subsection{Microscopic QRPA analysis for the wobbling motion}
\label{sec:QRPA_model}

A microscopic theory to treat the nuclear wobbling motion
in the small amplitude limit
is naturally provided by the QRPA
in the rotating frame,
or the self-consistent cranking plus QRPA~\cite{Mar79}.
Among the quadrupole tensors,
$Q_{ij}\propto x_i x_j - \delta_{ij} r^2/3$
($i,j=x,y,z$),
the negative signature operators of
$Q_y\equiv -Q_{zx}$ and $Q_z=iQ_{xy}$
are responsible for the wobbling motion.
Adopting the separable quadrupole interaction of
the form~(\ref{separable_interaction}),
the mean-field approximation simply replaces
one of the operators $QQ$ into its expectation value
$\mathcal{Q}(t)$, leading to the time-dependent mean field
\begin{equation}
h_{\rm UR}(t)=h_{\rm def} -\omega_{\rm rot} J_x 
- \kappa_y \mathcal{Q}_y(t) Q_y - \kappa_z \mathcal{Q}_z(t) Q_z ,
\label{UR}
\end{equation}
where $h_{\rm def}$ is a deformed single-particle Hamiltonian
that contains the fields associated with diagonal tensors,
$-\kappa_{ii} \mathcal{Q}_{ii} Q_{ii}$.
Although, in general, $h_{\rm def}$ is time-dependent,
we hereafter focus our discussion on the wobbling motion, and 
assume $h_{\rm def}$ is time-independent.
In this treatment of
equation~(\ref{UR}), the rotational axis stays along the $x$ axis
and the wobbling motion is represented by a fluctuation
in the orientation of deformed density distribution
induced by $\mathcal{Q}_y(t)$ and $\mathcal{Q}_z(t)$.
This picture corresponds to the uniformly rotating (UR) frame.

\begin{figure}[tb]
\centerline{
\includegraphics[clip,width=0.7\textwidth]{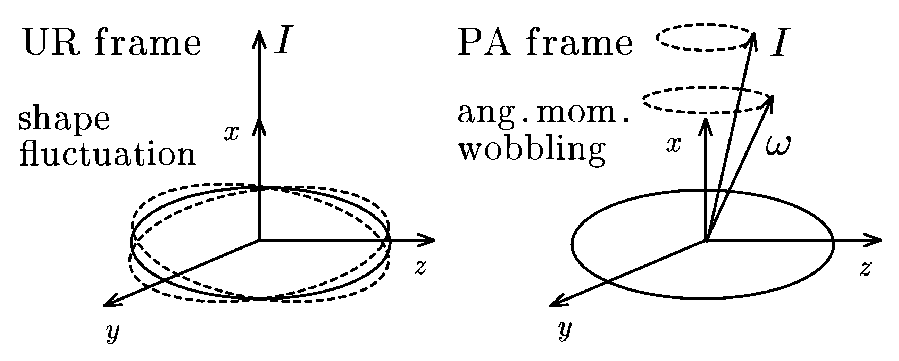}
}
\caption{
Two equivalent pictures of the wobbling motion:
one in the uniformly rotating (UR) frame (left)
and the other in the principal axis (PA) frame (right).
From reference \cite{SS06}.
}
\label{fig:UR_vs_PA}
\end{figure}

The small shape fluctuation induced by the off-diagonal quadrupole
tensors, $\mathcal{Q}_y$ and $\mathcal{Q}_z$, is not associated
with the real shape change from the equilibrium.
The same effect can be realized by rotating the reference frame to
the principal axis (PA) frame where the non-diagonal elements,
$Q_y$ and $Q_z$, of the quadrupole tensors vanish.
If we adopt this body-fixed frame,
the direction of the angular momentum fluctuates.
In the PA frame, since the rotation is no longer uniform,
the cranking model should be extended to a time-dependent one.
\begin{equation}
h_{\rm PA}(t)=h_{\rm def}-\vec{\omega}_{\rm rot}(t) \cdot \vec{J} .
\end{equation}
The rotor-model analysis of Bohr and Mottelson in section~\ref{sec:rotor_model}
has a direct connection to the PA picture.
In this picture, the frequency
$\vec{\omega}_{\rm rot}$
should be
treated as dynamical variables (operators).
In the small amplitude limit,
Marshalek proved the equivalence between the UR and the PA frames
and obtained the same expression~(\ref{wobbling_frequency}) for the
wobbling frequency,
with the moments of inertia calculated in the QRPA \cite{Mar79}.
It is generalized to arbitrary mean-field potentials and residual
interactions \cite{SS09}.
The two pictures are schematically illustrated in figure~\ref{fig:UR_vs_PA}.

The microscopic QRPA calculations were first performed with
the Nilsson potential and the separable quadrupole interactions
\cite{Mat90,SM95,MSM02,MSM04,SSM08}.
Later, it has been done with the Woods-Saxon potential and
an separable interaction which is determined by the symmetry
restoration condition \cite{SS09}.
In general, it is difficult to perform
the QRPA calculation for an odd-$A$ nucleus,
however, 
this is not a problem at a finite $\omega_{\rm rot}$
because the Kramers degeneracy is lifted and the RPA vacuum
is uniquely identified.
The calculated QRPA moments of inertia indicate a proper ordering
of moments of inertia,
$\mathcal{J}_x >\mathcal{J}_y >\mathcal{J}_z$,
for the wobbling mode in $^{163}$Lu.
Why is this ordering different from a naive expectation based
on the irrotational flow?

To answer the question, it is important to distinguish
the dynamic and the kinematic moments of inertia~\cite{BM81}.
The dynamic moment of inertia is defined by the second
derivative of the rotational energy,
$1/\mathcal{J}^{(2)}= d^2 E_I/dI^2$,
which is considered to be 
the one in the rotor model in equation~(\ref{triaxial_rotor}).
In contrast, the kinematic moment of inertia is defined by the first
derivative,
$1/\mathcal{J}^{(1)}= I^{-1} dE_I/dI =\omega_{\rm rot}/I$.
The largest moment of inertia, $\mathcal{J}_x$,
in the QRPA wobbling formalism of reference~\cite{Mar79} is the kinematic one,
more precisely,
$\mathcal{J}_x=\mathcal{J}^{(1)}_x \equiv \langle J_x \rangle/\omega_{\rm rot}$,
which is strongly influenced by
the alignment of the intrinsic angular momentum along the rotational ($x$) axis.
Generally the kinematic moment of inertia is larger than the dynamic one
because of the effect of alignment.
In $^{163}$Lu, the odd-proton quasiparticle mainly produces the alignment.
When the alignment is large enough, we could have
a rigid-body-like ordering,
$\mathcal{J}_x^{(1)} >\mathcal{J}_y >\mathcal{J}_z$,
for the positive $\gamma$ shape,
even if the dynamic moments of inertia are irrotational-like,
$\mathcal{J}_y >\mathcal{J}_x >\mathcal{J}_z$.
The QRPA moments of inertia automatically take into account
this effect.
Thus, {\it the alignment effect is crucial for the appearance of
the wobbling mode}, which was first pointed out in
references~\cite{MSM02,MSM04}.

The fact that the largest moment of inertia is the kinematic one can be
justified by the simple particle-rotor model as in reference~\cite{FD14}:
When the quasiparticle alignment $j$ is present, 
$J_x$ is replaced by $J_x-j$ in equation~(\ref{triaxial_rotor}).
Using $J_x =[{\vec{J}^2-(J_y^2+J_z^2)}]^{1/2}
\approx (I+\frac{1}{2})-(J_y^2+J_z^2)/(2I+1)$
which is valid for in high-spin limit $I \gg 1$,
we obtain
\begin{eqnarray}
&&\hspace*{-20mm}
H_{\scriptsize\mbox{p-rot}}=\frac{(J_x-j)^2}{2\mathcal{J}_x}
+\frac{J_y^2}{2\mathcal{J}_y}+\frac{J_z^2}{2\mathcal{J}_z}
\approx\frac{(I-j)(I-j+1)}{2\mathcal{J}_x}
\label{H_prot}
\\
&&\hspace*{-13mm}
+\left(\frac{1}{2\mathcal{J}_y}-\frac{1}{2\mathcal{J}_x}
 +\frac{1}{2\mathcal{J}_x}\frac{j}{I+\frac{1}{2}} \right)J_y^2
+\left(\frac{1}{2\mathcal{J}_z}-\frac{1}{2\mathcal{J}_x}
 +\frac{1}{2\mathcal{J}_x}\frac{j}{I+\frac{1}{2}} \right)J_z^2 .
\label{triaxial_protor}
\end{eqnarray}
Namely the inverse of the kinematic moment of inertia,
\begin{equation}
\frac{1}{\mathcal{J}_x^{(1)}}
 = \frac{\omega_{\rm rot}}{\langle J_x\rangle}
\approx \frac{1}{\mathcal{J}_x}\left( 1-\frac{j}{I+\frac{1}{2}}\right),
\label{J_x}
\end{equation}
appears in equation~(\ref{triaxial_protor}) in place of $1/\mathcal{J}_x$
in equation~(\ref{triaxial_rotor}).

The wobbling frequency (\ref{wobbling_frequency})
with $1/\mathcal{J}_x$ replaced by $1/\mathcal{J}_x^{(1)}$
of equation~(\ref{J_x})
first increases as spin increases
and then turns to decrease.
Thus the quasiparticle alignment also explains why the observed
wobbling frequency decreases as a function of $I$.
From equations~(\ref{triaxial_protor})
there is a critical angular momentum,
$I_c\equiv j (1-\mathcal{J}_x/\mathcal{J}_y )^{-1}-\frac{1}{2}$,
at which the wobbling frequency vanishes, $\omega_{\rm wob}=0$
(remember $\mathcal{J}_y>\mathcal{J}_x>\mathcal{J}_z$).
Beyond $I_c$, the wobbling mode ceases to exist,
because of the irrotational-like ordering,
$\mathcal{J}_y >\left. \mathcal{J}_x^{(1)}\right|_{I>I_c} >\mathcal{J}_z$.
In this way, for the case where the alignment
takes place along the axis of the intermediate dynamic moment of inertia,
the $I$-dependence of the original wobbling frequency
in section~\ref{sec:rotor_model} drastically changes.
Such a novel wobbling scheme was first pointed out in reference \cite{SMM04},
although the terms proportional to $j$ in equation~(\ref{triaxial_protor}),
i.e. the effect of alignment,
are interpreted as decreasing $\mathcal{J}_y$ and $\mathcal{J}_z$
instead of increasing $\mathcal{J}_x$.
The observed decreasing tendency of $\omega_{\rm wob}$ in the Lu isotopes
clearly suggests such a character.
In reference~\cite{FD14} it is called ``{\it transverse} wobbler''
in order to distinguish it from ``{\it longitudinal} wobbler''
where the quasiparticle aligns along the axis of the largest inertia,
$\mathcal{J}_x>\mathcal{J}_y>\mathcal{J}_z$.
In the longitudinal wobbler,
the frequency $\omega_{\rm wob}$ monotonically increases with $I$.
In fact, the microscopic QRPA calculations also predicted
the wobbling motion of increasing $\omega_{\rm wob}$ as a function of $I$
in nuclei of negative $\gamma$ shapes~\cite{Mat90,SM95},
in which the irrotational-type moments of inertia satisfy
the longitudinal condition,
$\mathcal{J}_x > \mathcal{J}_y > \mathcal{J}_z$.
A similar argument of the effect of alignment for the possible decrease
of $\omega_{\rm wob}$ has been discussed also in reference~\cite{HH03}.

It should be noticed that the three moments of inertia are assumed to be
independent of spin $I$ in the rotor model or the particle-rotor model.
In reality, however, the microscopically calculated QRPA moments of inertia
change
as functions of $I$, although their dependencies on $I$ are not so strong
in most cases~\cite{Mat90,SM95,MSM02,MSM04,SS09}.
One should take this into account in order to study precisely
how the wobbling frequency changes as a function of spin.
In reference~\cite{ST10}, introducing a rather strong spin-dependence
common to all three moments of inertia,
the decreasing tendency of $\omega_{\rm wob}$ is realized
in the particle-rotor coupling model with the inertia of
the rigid-body-like ordering, $\mathcal{J}_x >\mathcal{J}_y >\mathcal{J}_z$.

\begin{figure}[tb]
\centerline{
\includegraphics[clip,width=0.47\textwidth]{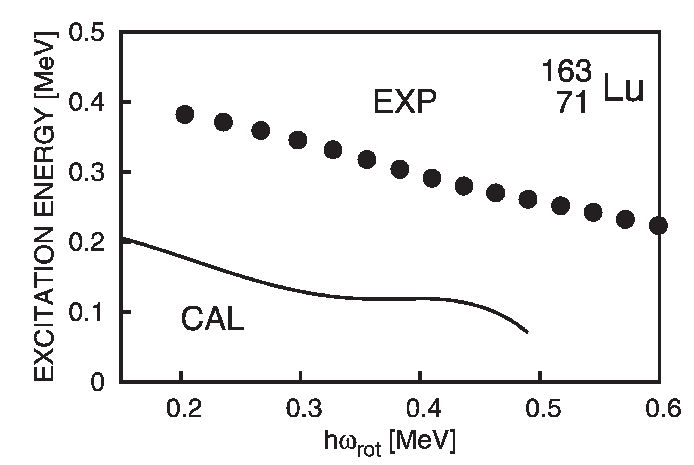}
\includegraphics[clip,width=0.47\textwidth]{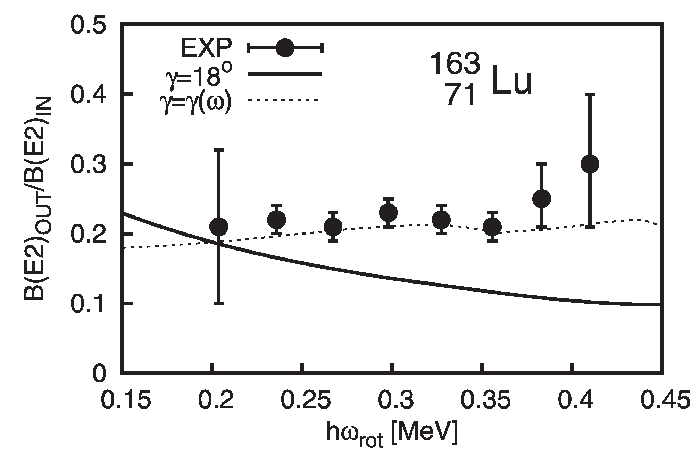}
}
\caption{
(Left)
Calculated and experimental wobbling frequencies as functions of the
rotational frequency.
(Right)
Calculated and experimental inter- to intraband $B(E2)$ ratio
as functions of the
rotational frequency.
See text for details.
Adapted from reference \cite{SS09}.
}
\label{fig:163Lu}
\end{figure}

In figure~\ref{fig:163Lu}, we show results of the QRPA calculation
based on the deformed Woods-Saxon potential~\cite{SS09}.
Note that there are no adjustable parameters in the calculation
because the minimal symmetry restoring interaction is employed,
which is uniquely fixed once the deformed mean-field is given.
The deformation parameters are determined by minimizing the TRS.
The calculated wobbling frequency has a proper trend,
though the absolute magnitude is underestimated by a few hundred keV.
The large interband $B(E2)$ values are rather well reproduced in the
calculation.
However, the observed ratio,
$B(E2;I\rightarrow I-1)_{\rm out}/B(E2;I\rightarrow I-2)_{\rm in}$,
seems to increase as a function of $I$,
while the calculated ratio decreases because of
the $1/I$ dependence of the interband transition.
The dotted line in the right panel of figure~\ref{fig:163Lu} indicates
the result obtained by artificially increasing the triaxiality
($\gamma$) at higher spins.
In fact, in order to explain the experimental $B(E2)$ ratios,
the triaxial parameters $\gamma({\rm den})\approx 20^\circ$
are necessary.
The $\gamma({\rm den})$ is defined with respect to the intrinsic quadrupole
moments calculated from the density distribution.
Here, it should be noted that the triaxial parameter $\gamma$ for the
potential shape is significantly different from $\gamma({\rm den})$ \cite{SSM08}
(See \ref{sec:triaxial_def} for details).
As far as we know, at present,
none of the microscopic calculations are able to
reproduce the triaxiality of $\gamma({\rm den})\approx 20^\circ$.
Another unsolved problem is that the observed $B(M1)$ values are
significantly overestimated.
In reference \cite{FD15},
the inclusion of the isovector separable orbital angular momentum
interaction is suggested to improve the agreement.
The nuclear wobbling motion has not been fully understood yet.

\subsection{Precession: Rotational band built on a high-$K$ isomer}
\label{sec:Precession}

In section~\ref{sec:QRPA_model}, we show that
the alignment of quasiparticle
is crucial for the wobbling motion
to appear in the Lu isotopes with the positive $\gamma$ shape.
An interesting extreme case
of the alignments is that
the nuclear shape is axially symmetric about the alignment ($x$) axis;
i.e. $\gamma=60^\circ$ (oblate) or $\gamma=-120^\circ$ (prolate)
in the Lund convention and the angular momentum
is supplied only by the alignments of quasiparticles.
It is expected that the optimal configurations of aligned quasiparticles
in the states of such shapes make the high-spin isomers, or the ``yrast traps'',
along the yrast line, see e.g. reference~\cite{VDS83}.
Although the rotational bands built
on the oblate isomers are not yet observed,
those on the prolate isomers have been well known~\cite{BM75}.
They are nothing but the high-$K$ rotational bands widely observed
in the Hf and W region, 
where many high-$j$ and high-$\Omega$ Nilsson
orbits are 
concentrated near the Fermi surface.
Here, $\Omega$ is the component of
single-particle angular momentum along the symmetry axis.

\begin{figure}[tb]
\centerline{
\includegraphics[clip,width=0.45\textwidth]{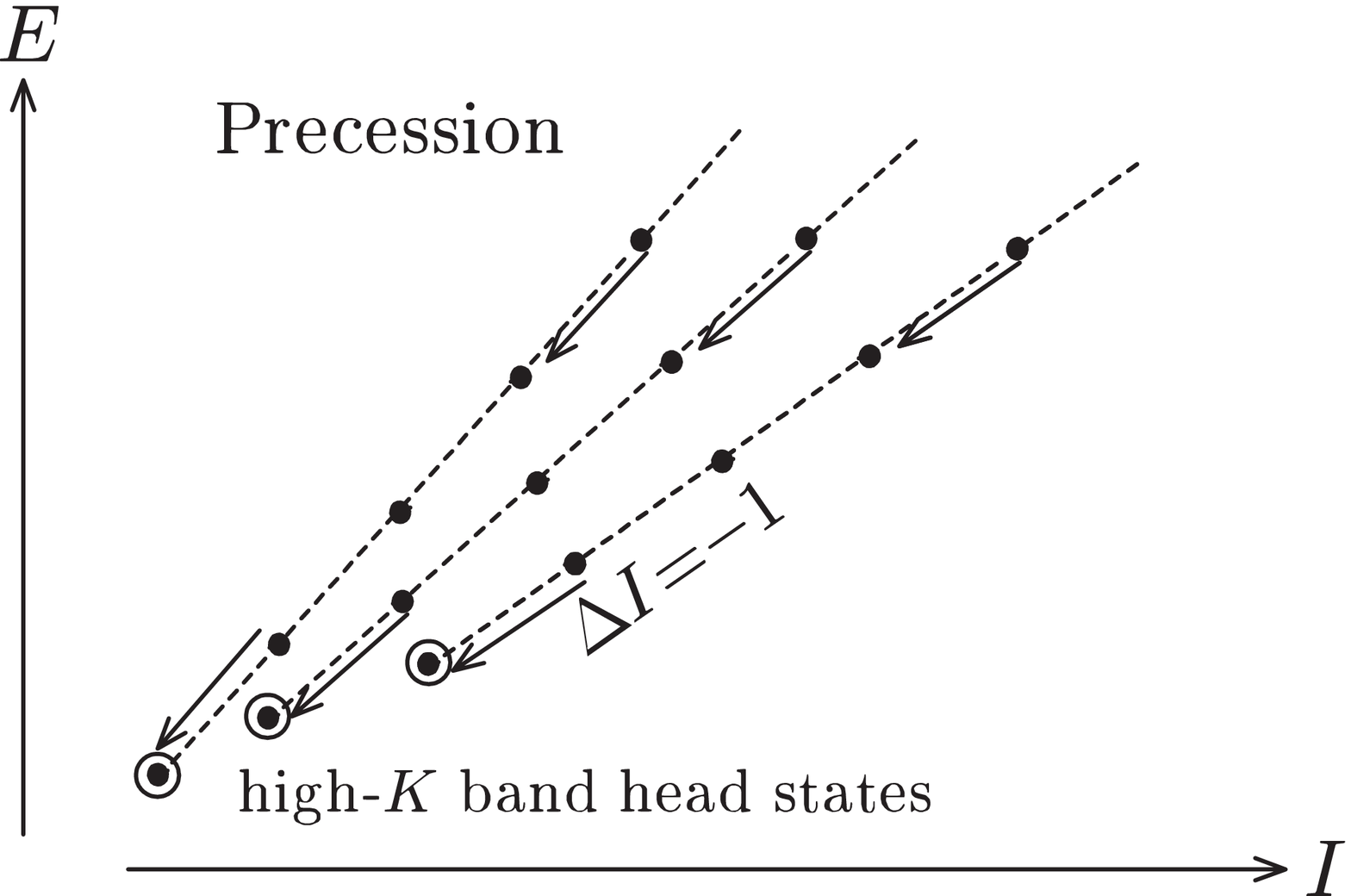}
\hspace*{0.05\textwidth}
\includegraphics[clip,width=0.18\textwidth]{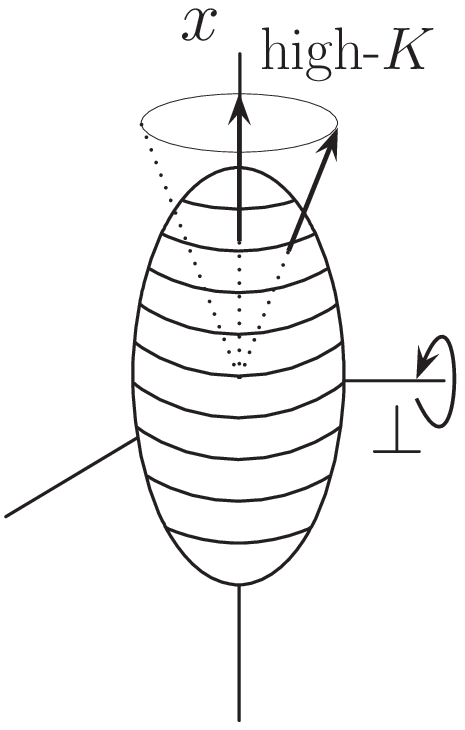}
}
\caption{
A schematic illustration of the precession motions.
(Left) Excitation spectra from reference~\cite{SMM05}.
There are no ${\Delta I}=2$ horizontal sequences
leaving only one ${\Delta I}=1$ vertical band for each high-$K$ state;
compare with figure~\ref{fig:wobbling}.
(Right) Superposition of the collective rotation about the perpendicular
axis makes the high-$K$ aligned angular momentum vector to precess around
the symmetry axis.
}
\label{fig:precession}
\end{figure}

We call this rotational motion ``precession'' because the aligned
angular momentum vector tilts like in the case of the wobbling motion
by superimposing the collective rotation about the perpendicular axis;
it is illustrated schematically in figure~\ref{fig:precession}.
Since the high-$K$ isomers have been known for many years,
they have been investigated by various methods;
e.g., by the cranked mean-field method~\cite{FP77},
by the RPA method based on
the sloping Fermi surface~\cite{Kur80,Kur82,AKLS81,Ska87},
or by the tiled axis cranking method~\cite{FNSW00}.
In reference~\cite{SMM05} it was considered as the axially symmetric limit
of the RPA wobbling formalism~\cite{Mar79} discussed in 
section \ref{sec:QRPA_model}.
In fact, the wobbling frequency in equation~(\ref{wobbling_frequency}) becomes
$\omega_{\rm wob}=I/{\cal J}_\bot - \omega_{\rm rot}$, where
the perpendicular moment of inertia is denoted as
${\cal J}_\bot\equiv {\cal J}_y={\cal J}_z$ in the axially symmetric
limit and the rotational frequency about the main rotation axis
is $\omega_{\rm rot}=I/{\cal J}^{(1)}_x$.
Here, the dynamic moment of inertia (${\cal J}_x\rightarrow 0$) is
replaced by the kinematic inertia ${\cal J}^{(1)}_x$
with the aligned angular momentum $I_x=K$ (we here use $K$ in place of $j$).
On the other hand the rotor Hamiltonian~(\ref{triaxial_rotor}) in this case
reduces to $H_{\rm rot}=(I_y^2+I_z^2)/(2{\cal J}_\bot)$
so that the energy spectrum is given simply by
$E_{{\scriptsize\mbox{high-}}K}=[I(I+1)-K^2]/(2{\cal J}_\bot)$,
which can be rewritten as
\begin{equation}
 E_{{\scriptsize\mbox{high-}}K}=
 \omega_{\rm prec}\Bigl(n+\frac{1}{2}+\frac{n(n+1)}{2K}\Bigr) ,
\label{eq:prec}
\end{equation}
introducing the precession phonon number $n\equiv I-K$.
Here the precession frequency is given by
$\omega_{\rm prec}\equiv K/{\cal J}_\bot
=[\omega_{\rm wob}+\omega_{\rm rot}]_{I=K}$.
Thus the precession motion can be described by the harmonic excitation
of $n$-phonons as long as $n \ll K$.  The difference of frequencies
$\omega_{\rm prec}$ from $\omega_{\rm wob}$ is due to the fact that
the wobbling motion is treated in the body-fixed frame while the precession
motion in the laboratory-frame.
Remember the transformation of the energy
into the routhian in the rotating frame
in section~\ref{sec:cranking_at_high_spin}, $E'=E-\omega_{\rm rot}I_x$,
and that the precession mode transfer the angular momentum by one unit
$\Delta I_x=1$.
Since there is no collective rotation about $x$ axis, 
the rotational frequency $\omega_{\rm rot}$ is a redundant variable
and any observable quantities do not depend on it;
$\omega_{\rm wob}$ does depend while $\omega_{\rm prec}$ does not.
As it is discussed in reference~\cite{SMM05} not only the energy but also
the electromagnetic transitions, like $E2$ and $M1$,
can be treated with the multi-phonon picture
as long as the phonon number $n$ is much smaller than $K$.

\begin{figure}[tb]
\centerline{
\includegraphics[clip,width=0.8\textwidth]{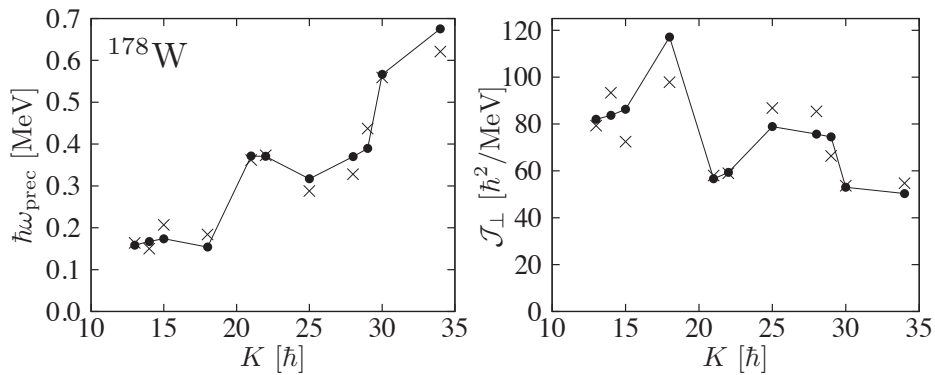}
}
\caption{
(Left) 
Calculated (filled circles) and experimental (crosses)
precession frequencies $\omega_{\rm prec}$
for various high-$K$ isomers in $^{178}$W,
where the experimental data is denoted by crosses.
(Right) 
Calculated (filled circles) and experimental (crosses)
moments of inertia perpendicular to the symmetry axis,
estimated by ${\cal J}_\bot=K/\omega_{\rm prec}$.
Based on the result of reference~\cite{Sho09}.
}
\label{fig:precRPAws}
\end{figure}

We show in figure~\ref{fig:precRPAws} the result of precession frequencies
for a number of $K$ isomers in $^{178}$W calculated
by using the axially symmetric limit
of the Woods-Saxon QRPA wobbling formalism~\cite{SS09}.
Compared with the corresponding calculation of reference~\cite{SMM05},
in which the Nilsson mean-field potential is employed,
considerable improvement can be seen and a good agreement
with experimental data is obtained.
In the right panel of figure~\ref{fig:precRPAws},
the estimated moments of inertia ${\cal J}_\bot=K/\omega_{\rm prec}$
are also shown.
The agreement is much better than the simple mean-field calculation,
e.g.~\cite{FP77},
because the effect of residual interaction is taken into account in the QRPA.
It can be seen that the moments of inertia
take various values depending on the isomer configurations.
They are considerably larger than the moment of inertia of
the ground state rotational band estimated from the first $2^+$ state,
$\mathcal{J}_{\rm gr} \approx 28$ [$\hbar^2$/MeV].
They do not show a simple correlation with the $K$ quantum number,
and do not approach to 
the rigid-body value (with $\epsilon_2=0.240$).
${\cal J}_{\rm rig} \approx 88$ [$\hbar^2$/MeV],
even at considerably high spin.
Their properties strongly depend on what kind of quasiparticles
contribute to those high-$K$ isomers
in which the numbers of quasiparticles are from four to ten.
See reference~\cite{SMM05} for precise configuration assignments.
At an extreme high spin, we can even
imagine possible existence of
torus-shape isomers and their precession motions \cite{IMMI14}.

\section{Summary}

Bohr and Mottelson have explored a variety of fields in the nuclear
structure physics.
Among them, we have discussed selected topics related to
the nuclear deformation and rotation.
First, we presented  the concept of the
symmetry breaking in the unified model.
The symmetry broken state is not stable in finite systems, such as nuclei.
The correlation time induced by the quantum fluctuation
is a key to understand
the interplay between the symmetry breaking and restoration.
The finite-size effect associated with the zero-point motion may 
hinder the symmetry breaking.

The coupling between intrinsic and rotational motions is
well described by the cranking model.
Since the model assumes a semiclassical treatment of the
nuclear rotation (angular momentum),
the model requires the quantization in the low-spin limit.
We show a possible quantization of the cranking model,
which is applicable to calculation of transition matrix elements
at low spin.
This can be regarded as a kind of hybrid model of
the unified model and the cranking model.
It is applied to electromagnetic decay properties
of vibrational bands and high-$K$ isomers.

In the high-spin region, the cranking model is a golden tool to study
the nuclear structure under a strong Coriolis and centrifugal
field.
We discussed effects of the rapid rotation on the octupole vibrations,
which are nicely treated with the QRPA
in the uniformly rotating frame (one-dimensional cranking).
The calculation reproduces the experimental data, showing the
phonon alignment and loss of collectivity (phonon breakdown).

The closed shell configurations of the superdeformed (SD) states
are characterized by the $2:2:1$ shell structure.
This shell structure has the $K^\pi=1^-$ octupole mode as a soft mode.
Away from the SD magic numbers with many valence nucleons,
the prolate SD nucleus could show instability toward a bending banana shape.

The triaxial deformation produces the three dimensional
non-uniform rotation, which is called wobbling motion.
The QRPA in the uniformly rotating frame provides a microscopic
tool to calculate the wobbling and precession
modes of excitation.
The experimental data are qualitatively reproduced.
This microscopic study
clearly indicates the importance of the quasiparticle alignment
for the existence of the wobbling mode.
The self-consistently calculated triaxial deformation seems
to be smaller than what experimental data indicate,
which is an important open problem.

The nucleus provides a wonderful opportunity to study a finite
system going through many kinds of symmetry breaking,
under a variety of extreme circumstances, such as
large angular momentum, deformation, and isospin.
The topics we have discussed in this paper were pioneered by
Bohr and Mottelson who gave us a deep insight into
nuclear structure and quantum many-body physics.
There are still many open issues in these fields
which are waiting for future studies.

\ack

This work was supported in part by JSPS KAKENHI Grant
No. 25287065 and
by Interdisciplinary Computational Science Program in CCS,
University of Tsukuba.

\appendix

\section{Remarks on the triaxial deformation}
\label{sec:triaxial_def}

The values of the triaxiality parameter $\gamma$ can be significantly
different depending on their definitions.
This was first pointed out
in the Appendix B of reference~\cite{SM84} and more recently discussed again
in relation to the wobbling motion in reference~\cite{SSM08}.
The most basic definition is
$\gamma(\mbox{den})\equiv
-\tan^{-1}(\langle Q_{22}\rangle/\langle Q_{20}\rangle)$
by using the intrinsic quadrupole moments,
which is directly related to
the $E2$ transition probability.
In phenomenological potential models,
such as the Nilsson and the Woods-Saxon potentials,
the triaxial deformation $\gamma\equiv\gamma(\mbox{pot})$ is introduced
to define the shape of the potential.
For example, it is defined based on the stretched coordinate
in the Nilsson model, $\gamma(\mbox{pot:Nils})$,
and on the radius parametrization
$R(\theta,\phi)\propto (1+\sum a_{\lambda\mu}Y_{\lambda\mu})$
in the Woods-Saxon model, $\gamma(\mbox{pot:WS})$.

With the uniform density assumption,
the triaxiality parameter $\gamma(\mbox{geo})$ can be calculated
in the same way as $\gamma(\mbox{den})$ for a given $\gamma(\mbox{pot})$.
Then, $\gamma(\mbox{den})\approx \gamma(\mbox{geo})$ holds
in a good approximation
near the self-consistent point~\cite{SSM08},
reflecting the short-range nature
of the nucleon-nucleon interaction.
However, it should be noted that
$\gamma(\mbox{geo})$ is different from $\gamma(\mbox{pot})$,
see e.g. reference~\cite{HM88}.
These different definitions are sometimes confused.

Here we would like to take another well-known example
for the harmonic oscillator model with the quadrupole deformed potential,
$-(\omega/b^2)\beta[\cos\gamma\,Q_{20}-\sin\gamma(Q_{22}+Q_{2-2})/\sqrt{2}]$,
where $b$ is the oscillator length parameter for the frequency $\omega$,
and we call these $\beta$ and $\gamma$ as the \mbox{(pot:HO)}-parametrization.
Then the deformed shape of potential is the ellipsoid
with the anisotropic frequencies
$\omega_k^2=\omega^2 [1-\sqrt{\frac{5}{4\pi}}\beta\cos(\gamma+\frac{2\pi}{3}k)]$
($k=1,2,3=x,y,z$).  As is discussed in section~\ref{sec:self-consistency},
the isotropic velocity distribution condition
in equation~(\ref{self_consistency_ho}) leads to
$\langle x^2_k \rangle \propto \omega_k^{-2}$ and then
$\gamma(\mbox{den})=\gamma(\mbox{geo})
=\tan^{-1}[\sqrt{3}(\omega_y^{-2}-\omega_x^{-2})
/(2\omega_z^{-2}-\omega_x^{-2}-\omega_y^{-2})]
\approx (1-3\sqrt{\frac{5}{4\pi}}\beta)\gamma$ for $\beta,|\gamma| \ll 1$,
which shows that the $\gamma(\mbox{geo})$ can be very different
from $\gamma\equiv\gamma(\mbox{pot:HO})$
for larger $\beta\equiv\beta(\mbox{pot:HO})$ values.
Similar differences are pointed out for
the Nilsson and the Woods-Saxon potentials in reference~\cite{SSM08}.
Introducing $\epsilon=\sqrt{\frac{45}{16\pi}}\beta$ in place of $\beta$,
we have
\begin{equation}
\fl
\gamma(\mbox{geo})\approx (1-2\epsilon)\gamma(\mbox{pot:HO})
\approx (1-\frac{3}{2}\epsilon)\gamma(\mbox{pot:Nils})
\approx (1-\frac{8}{7}\epsilon)\gamma(\mbox{pot:WS}) .
\end{equation}
Strictly speaking, the $\epsilon$ is also different in each definition,
but that is neglected here.
The difference between $\gamma(\mbox{geo})$ and $\gamma(\mbox{pot})$
is largest for the \mbox{(pot:HO)}-parametrization among these three examples.

The deformed shape for the wobbling motion
in the Lu isotopes is predicted to be
$\epsilon \sim 0.4$ and $\gamma\sim 20^\circ$~\cite{Gor04}
in the \mbox{(pot:Nils)}-parametrization.
Assuming purely ellipsoidal shape it leads to
$\gamma(\mbox{geo})\approx 11^\circ$~\cite{SSM08},
which is significantly smaller than the value $20^\circ$.
The experimentally measured $B(E2)$ values seem to indicate
$\gamma(\mbox{den})\approx 20^\circ$ or even larger values
at higher spins~\cite{SS09}.   In order to obtain
$\gamma(\mbox{geo})\approx \gamma(\mbox{den})\approx 20^\circ$,
$\gamma(\mbox{pot:Nils})\approx 31^\circ$
and
$\gamma(\mbox{pot:HO})\approx 36^\circ$
are necessary with keeping $\epsilon(\mbox{pot:Nils})\approx 0.4$.
In the same way, assuming only the quadrupole deformation,
$\gamma(\mbox{pot:WS})\approx 28^\circ$ is necessary
with $\beta(\mbox{pot:WS})\approx 0.4\times\sqrt{\frac{16\pi}{45}}$.
Thus the required values of the potential triaxiality parameters
are considerably larger, which are not obtained in any TRS calculations.
The situation is the same for the self-consistent mean-field calculation.
For example, we obtain $\gamma(\mbox{den})\approx 11^\circ$
for $^{163}$Lu by the cranked HFB calculation
with the Gogny D1S force~\cite{Shi16}, which is consistent
with the Nilsson TRS calculations.
None of the microscopic calculations are able to
reproduce the triaxiality of $\gamma(\mbox{den})\approx 20^\circ$,
see also reference~\cite{FD15}.

\section*{References}
\bibliographystyle{iopart-num}
\bibliography{nuclear_physics,chemical_physics,myself,current}

\end{document}